\documentclass{optica-article}
\usepackage[latin1]{inputenc}
\usepackage{amsmath,bm}
\usepackage{amsmath}
\usepackage{amsfonts}
\usepackage{newtxmath}
\usepackage{newtxtext}
\usepackage{float}
\usepackage{cite}
\usepackage{breqn}
\usepackage{graphicx}
\usepackage{subcaption}
\usepackage{subfig}
\usepackage{cite}
\usepackage{comment}
\usepackage{siunitx}

\journal{optcon}

\articletype{Research Article}

\usepackage{lineno}
\makeatletter
\providecommand\sf@counterlist{}
\makeatother

\begin{document}

\title{Advanced Knife-Edge free Self-Aligned Colour Schlieren Imaging with Extended Measuring Range}

\author{Shubham Saxena\authormark{1}, and Xu Wang\authormark{2*}}

\address{\authormark{1,2}Institute of Physics and Quantum Science, Heriot-Watt University, Edinburgh EH14 4AS, United Kingdom\\}

\email{\authormark{1}ss2330@hw.ac.uk}
\email{\authormark{2} *corresponding author, x.wang@hw.ac.uk}


\begin{abstract}
Schlieren imaging is a powerful, non-intrusive method widely used to visualize refractive index gradients in fluid dynamics and heat transfer studies, essential in fields like aerospace engineering, combustion analysis, and supersonic flow visualization. Traditional Schlieren imaging setups employ a knife-edge cut-off to differentiate between deviated and non-deviated light rays, creating contrast. However, this approach requires highly precise alignment and limits both usability and measurement range, restricting accurate observation of high-gradient phenomena such as combustion processes at elevated temperatures. Here, we proposed a simplified Schlieren imaging method that eliminates the knife edge by introducing a novel integration of an extended RGB segmented colour source with a circular aperture. Our approach significantly simplifies alignment, enables colour visualization, and we achieve expanded measuring range by approximately threefold. Experimental validation, involving imaging hot air plumes generated from butane combustion and transparent adhesive tape tests, demonstrate significant improved performance, including enhanced visibility near regions of high temperature gradients, improved edge detection, and precise visualization of large refractive gradients. This simplified yet robust Schlieren imaging approach substantially broadens the scope and ease of application, facilitating detailed investigations of previously challenging fluid phenomena.
\end{abstract}

\section{Introduction}

Schlieren imaging, developed by August Toepler, from the early 19th century, is a well-known optical method used to observe variations in refractive index within transparent objects and has been one of the key methods for visualizing refractive index gradients \cite{settles2001schlieren, schardin2007schlierenverfahren}. This technique is widely used for qualitative visual analysis and is renowned for its straightforward methodology, involving simple optical components and easy setup suitable for a wide range of applications \cite{settles2017review, elsinga2004assessment, stanic1978quantitative, agrawal1998three, tregub2004color, alvarez2009temperature}. Schlieren systems have historically been used to observe a wide range of phenomena, such as the striations in blown glass, animal and human respiratory fluxes to precisely detect disease transmission, airplane shock waves, heat emissions from various sources, plasma flow imaging, and range of particle propagation in a flow \cite{barnes1945schlieren, xu2017human, weinstein1996electronic, panigrahi2012schlieren, tang2011observing, tang2009schlieren, tang2013airflow, estevadeordal2004schlieren}. This non-intrusive optical approach is especially useful in fluid dynamics research because of its high sensitivity to fluid density arising from temperature, pressure, or composition variations and its non-interfering nature with the flow \cite{curtis2015schlieren}. In the 1930s, Schmidt \cite{schmidt1932schlierenaufnahmen} and Schardin \cite{schardin1934toeplersche, schardin2007schlierenverfahren} proposed that it might be used to measure variations in refractive index, from which density and temperature in fluid flow scenarios can be derived. Consequently, various methods have been developed to quantify Schlieren images \cite{schmidt1932schlierenaufnahmen, schardin1934toeplersche, elsinga2004assessment}.

Schlieren imaging came to a standstill in the 20th century, despite its early triumphs. The complex optical alignment, immobility, and measurement limitations were the main causes of this situation \cite{settles2001schlieren}. Recent advances in Schlieren imaging, driven by improvements in digital cameras and computational power, have led to significant growth in the field \cite{settles2017review, harvey2018visualization}. However, the inherent limitations remain in getting a large measurement range of the system within these systems. Therefore, it faced various issues across different fields, such as limitations in measuring high temperatures \cite{barnes1945schlieren, alvarez2015proper}. Previous studies have attempted to extend the measurement range of Schlieren systems by capturing multiple images at different exposure times using a monochrome camera and combining them through high dynamic range (HDR) techniques \cite{martinez2016wide}. While effective in increasing the dynamic range for static or quasi-steady thermal fields, such methods are not suitable for dynamic scenes due to their multi-frame nature. In contrast, our approach provides an extended measurement range in a single shot, enabling real-time visualization without requiring scene stability or complex image fusion.

\begin{figure}[H]
    \centering
    \begin{subfigure}[b]{0.45\textheight}
        \centering
        \includegraphics[width=\textwidth]{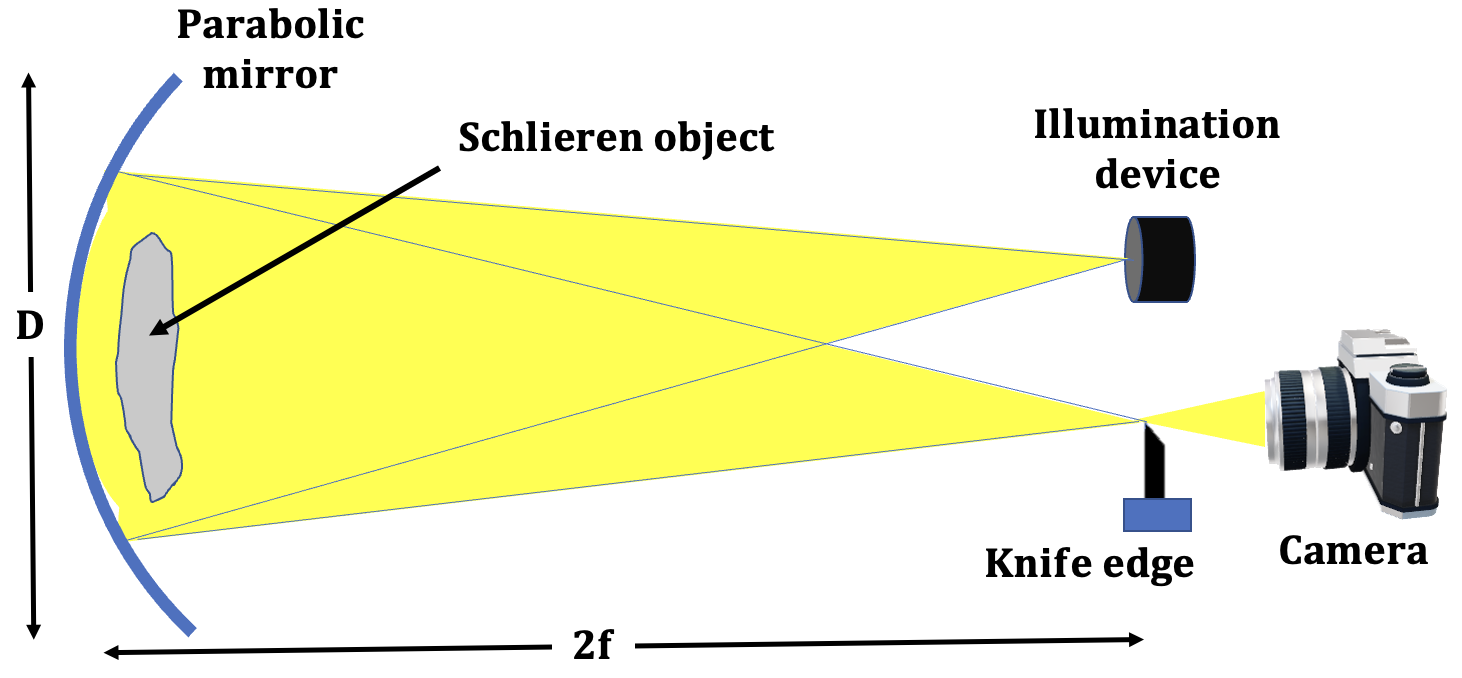}
        \caption{System setup}
        \label{tradknife1a}
    \end{subfigure}
    \hfill
    \begin{subfigure}[b]{0.27\textwidth}
        \centering
        \includegraphics[width=\textwidth]{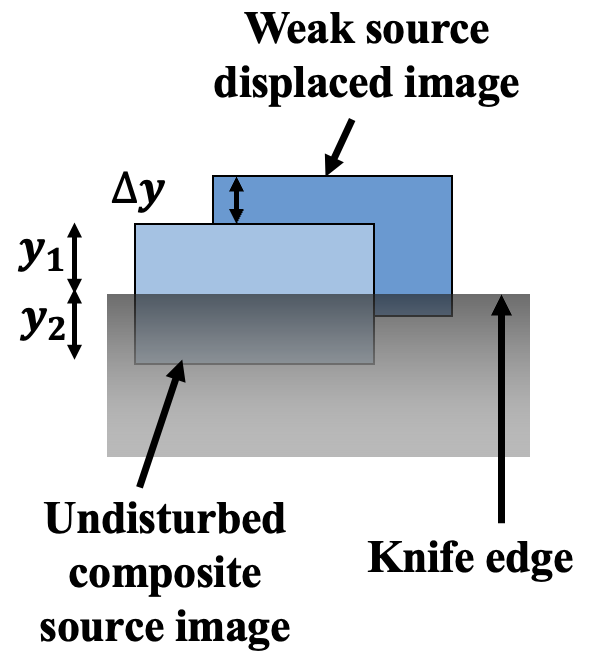}
        \caption{Knife edge plane}
         \label{tradknife1b}
    \end{subfigure}
    \caption{The single mirror Schlieren system.}
    \label{tradknife1}
\end{figure}

In a traditional single-mirror Schlieren setup, a point-like light source, a concave or parabolic mirror, a knife edge, and an imaging device are used, as illustrated in Fig.~\ref{tradknife1a}. Refractive index gradients introduced by the Schlieren object deflect light rays from their original paths.
The system differentiates between deflected and undeflected rays, translating these deviations into intensity contrast for image formation.
To differentiate them, a point light source is positioned off-axis at a distance equal to the mirror's radius of curvature. The mirror forms a real image of the source in the same plane, near its original position. A knife edge is then placed at this position to partially obstruct the light, converting the angular deflections caused by these gradients into image contrast.
An imaging device is positioned behind the knife edge to record the Schlieren image.
Each point within the Schlieren field causes a unique deflection, resulting in a corresponding shift in the light source image at the knife edge, which introduces contrast by selectively blocking portions of the angularly deflected light rays. Consequently, varying light intensities, reduced or increased relative to the background, reach the camera, visualizing the otherwise invisible structures in the flow field.
Therefore, the range of measurable angular deviations mapped onto the available pixel intensity range is determined by the size of the projected light source image at the knife edge. A smaller source size compresses the intensity transition over a narrower angular range, while a larger source distributes the same intensity variation across a broader range, enabling observation of larger refractive-index gradients.

\vspace{-0.2cm}

In Fig.~\ref{tradknife1b}, contrast continues to vary as the source image shifts upward, until \(\Delta y\) reaches \(y_2\). Similarly, for a downward shift, the variation persists until \(\Delta y\) reaches \(-y_1\).
Beyond this limit, contrast no longer changes, resulting in loss of detail for strong gradients that cause shifts exceeding \(y_1\) (downwards) or \(y_2\) (upwards). 
The system's sensitivity to maximum angular deviations defines its measuring range, corresponding to the refraction angles that shift the light source image to \(y_1\) and \(y_2\) at the knife-edge plane in their direction.
A universally recommended half cut-off for equal range in both directions, is broadly used in Schlieren experiments \cite{settles2001schlieren}.

The requirement for a point-like light source imposes inherent limitations on the system's measuring range, thereby reducing its capability to detect large light deviations caused by strong refractive index gradients, such as those arising from high temperature or pressure variations in the medium \cite{settles2001schlieren}.
A large measuring range is essential in scenarios involving strong refractive index gradients, such as in combustion processes~\cite{vanselow2019particle, vetrano2005assessment}. This range is directly proportional to the size of the projected light source image and can be increased by extending the limits \(y_1\) and \(y_2\) at the knife-edge plane.
A large measuring range is essential in scenarios involving strong refractive index gradients, such as in combustion processes \cite{vanselow2019particle, vetrano2005assessment}. This range is directly proportional to the size of the projected light source image and can be increased by expanding the limits \(y_1\) and \(y_2\) at the knife-edge plane.
However, this solution introduces a trade-off with image sensitivity. A \SI{50}{\percent} cutoff provides limited sensitivity for broad light sources; a very narrow source is crucial for achieving high sensitivity, highlighting an inherent conflict between maximizing sensitivity and expanding the measuring range \cite{settles2001schlieren}.
Using a broader light source improves the measurable range, as larger angular deviations are required to shift the source image fully on or off the knife edge. However, this simultaneously reduces sensitivity in the captured Schlieren images.
Increasing the cutoff to around 90\% may enhance sensitivity, but it leads to an asymmetric measuring range, broader on the brightening side of the image and narrower on the darkening side \cite{settles2001schlieren}.

A camera's dynamic range is fundamentally constrained by the number of grey levels it can capture. For instance, an 8-bit camera can represent only 256 discrete intensity values. As a result, continuous variations in illumination must be rounded to the nearest available grey level, leading to quantization errors and a reduction in sensitivity. This limitation becomes more pronounced as the measurement range increases, since more intermediate intensity values fall between the fixed grey levels and cannot be accurately resolved. Consequently, fine details may be lost, reducing both contrast and the precision of quantitative measurements \cite{zhang2018quantative}.

Furthermore, knife edge in traditional Schlieren system introduces a strong dependence of sensitivity on the cutoff fraction, making the system highly sensitive to small alignment errors and resulting in potential non-linearity in contrast response. It affect the accuracy of quantitative measurements such as temperature and density fields. In contrast, circular apertures offer more robust behavior and are particularly beneficial when high spatial resolution is required, as noted by Schmidt-Bleker et al. in their analysis of Schlieren contrast linearization~\cite{schmidt2015quantitative}.

\section{Methods}
Our approach increases the number of measurable intensity levels through a broader light source, enabling both preservation of sensitivity and extended measurement range.
In our system, instead of a uniform small light source, we employ a segmented source design that enables the camera to resolve individual grey levels from each segment separately. This approach preserves sensitivity by encoding a wider range of deflection-induced intensity variations within the same field of view.
For a broad light source with \(n\) distinct segments (e.g., different colours or wavelengths), the system can resolve up to \(n\) times as many intensity levels compared to the traditional approach. For instance, with an 8-bit camera, instead of representing \(255\) grey levels, the use of \(n\) segments enables measurement of up to \(255\times n\) intensity levels.
Therefore, while maintaining the same sensitivity, the light source dimensions can be increased by a factor of \(n\) compared to a traditional setup. The effective number of intensity levels, \(255\times n\), ensures a consistent mapping of grey values relative to deflection angles in both conventional and segmented systems, by keeping the intensity step size unchanged. These additional levels are distributed over a broader range of angular deflections. As a result, the system preserves sensitivity while increasing the measuring range determined by the size of the light source image, by a factor of \(n\). The measuring range of the new system, 
\(\epsilon_{max}^{(N)}\), is 
\(n\) times that of the traditional system, 
\(\epsilon_{max}^{(a)}\), as shown in Eq. \ref{eq1mr}.

\begin{equation}
    \epsilon_{\text{max}}^{(N)} = n\, \epsilon_{\text{max}}^{(a)}
    \label{eq1mr}
\end{equation}

Furthermore, the cutoff must be applied such that each shift of the light source image on the cutoff plane corresponds to a unique combination of grey values from the individual segments, yielding a total of \(255 \times n\) possible intensity levels. Unlike the traditional small uniform light source, the segmented configuration in the proposed system makes a knife edge less suitable as a cutoff. In contrast, a circular aperture is more appropriate for two key reasons: (1) for a specific deviation, it allows light from only a single or at most two partial segments to enter the camera, simplifying the definition of unique segment combinations for the calibration curves; and (2) modern imaging devices (e.g., Digital Single-Lens Reflex camera or mobile phone cameras) typically incorporate an internal circular aperture, which can be directly utilized as the cutoff. This enables the implementation of an external-cutoff-free, self-aligned Schlieren system.
One suitable light source configuration involves utilising the RGB colour channels of a colour CMOS display to generate distinct RGB colour strips. In the following sections, we present the experimental implementation of the proposed system and compare its performance with that of traditional Schlieren imaging methods.

A suitable light source design involves using RGB colour strips, which can be effectively distinguished by a colour CMOS camera through its separate red, green, and blue channels. In the following sections, we present the experimental validation of the proposed system and compare its performance with traditional Schlieren imaging methods.

\subsection{System description}
In the experiment, we constructed the proposed system using a custom light source, a \(200\,\text{mm}\) (8-inch) parabolic mirror with a \(1.2\,\text{m}\) focal length, a circular aperture, and an imaging device (e.g. a CMOS camera or a mobile phone camera), as shown in Fig.~\ref{fig:kefa1a}.
In this setup, the point light source is replaced by a proposed segmented light source, and the knife edge is substituted with a circular aperture. The aperture can be implemented either as an external cut-off element or as the internal aperture of the camera. Using the internal aperture simplifies the system significantly, whereas an external aperture allows for more precise control and adjustability.

The pivotal element of this system is a segmented RGB light source, composed of repeated sets of four vertical segments in the sequence: red, blue, green, and black. The projected image of this repeated pattern aligns naturally with the aperture, making the system self-aligned. The pattern can be generated either digitally or through a printed mask illuminated by white light.
Segment's width is selected based on the camera's aperture size, estimated from the desired measuring range for imaging the specific Schlieren object.
Depending on the segment's position relative to the aperture, the resulting Schlieren image background appears as one of six distinct shades: Red, Blue, Green, Red-Blue, Blue-Green, or Black.
Each background can appear in a range of shades depending on the alignment, and for an 8-bit camera, up to \(255\) distinct intensity levels are available for each background.

The segmentation allows precise control over the background colour in the Schlieren image, which can be adjusted by simply shifting the light source within its plane without modifying other system components, including the cut-off. This flexibility is particularly advantageous for achieving optimal colour contrast in qualitative imaging. By selecting an appropriate background, such as black, the system can enhance the visibility of fine-flow features and subtle refractive index variations.

\subsection{Experimental setup}
\begin{figure}[H]
    \centering\includegraphics[width=12cm]{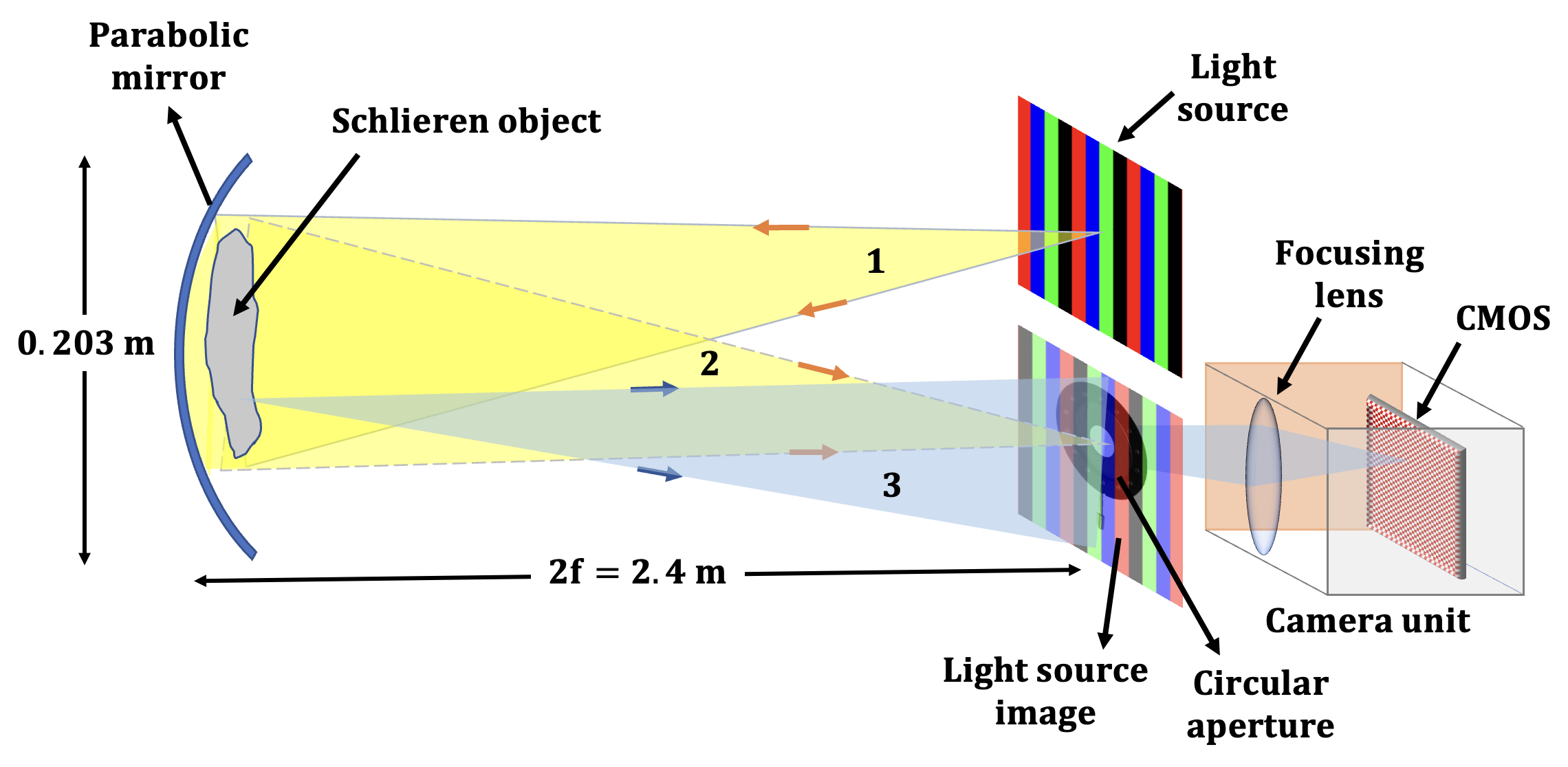}
    \caption{Experimental setup of the proposed knife-edge-free off-axis Schlieren system using the new light source.}
    \label{fig:kefa1a}
\end{figure}

Fig.~\ref{fig:kefa1a} illustrates the experimental setup of the proposed knife-edge-free off-axis Schlieren imaging system, employing the segmented light source. The Schlieren image formation involves two separate imaging paths: the parabolic mirror images the light source onto the aperture plane, while the camera lens is focused on the Schlieren object. Accurate source imaging is a critical first step, as it ensures necessary illumination of the object. 
Following refraction through the object (as governed by Snell's law) and selective cut-off at the aperture, the resulting light distribution generates contrast in the captured Schlieren image.
\vspace{-0.3cm}

The segmented light source is positioned at the mirror's radius of curvature and slightly offset from its geometric axis within the same vertical plane.
The light rays emerge from the segmented source; as illustrated in Fig.~\ref{fig:kefa1a}, beam 1 originates from a point on the source, reflects off the parabolic mirror, and is converged as beam 2, forming its image at a distance \(2f\), offset from the original point.
The full light source image is formed at the focal plane, where a circular aperture is placed to spatially modulate the light entering the imaging system.
The aperture diameter is matched to the width of a single vertical colour segment in the source image. As a result, even a slight horizontal shift of the light source image at the aperture plane alters which segments align with the open region of the aperture, thereby changing the amount and spectral composition of light passing through it.

The camera unit is positioned after the circular aperture and focused on the Schlieren object. Light beam 3, originating from a point on the object located near the mirror, forms a weak image of the segmented source at the aperture plane. This image is laterally shifted from the undeviated source image, with the magnitude of the shift determined by the local refractive index gradient at that object point. Due to this shift, a different combination of segments aligns to the open area of the aperture compared to the background, resulting in a modified light spectrum reaching the camera. This light is then focused onto the corresponding pixel of the sensor, forming the image of that object element. In the final Schlieren image, each pixel encodes a unique RGB combination based on the shift in the weak source image. The background colour is defined by the segment combination transmitted from undeviated rays.

\subsection{Geometrical Optics Model}
In Schlieren imaging, the output is a two-dimensional image composed of spatially varying intensity values across the \(x\) and \(y\) axes as an array of image elements. In an RGB image, each pixel encodes colour information through varying combinations of red, green, and blue components, which can be expressed as separate grayscale intensities for each colour channel.

\begin{figure}[H]
    \centering
    \begin{subfigure}[b]{0.333\textheight}
        \centering
        \includegraphics[width=\textwidth]{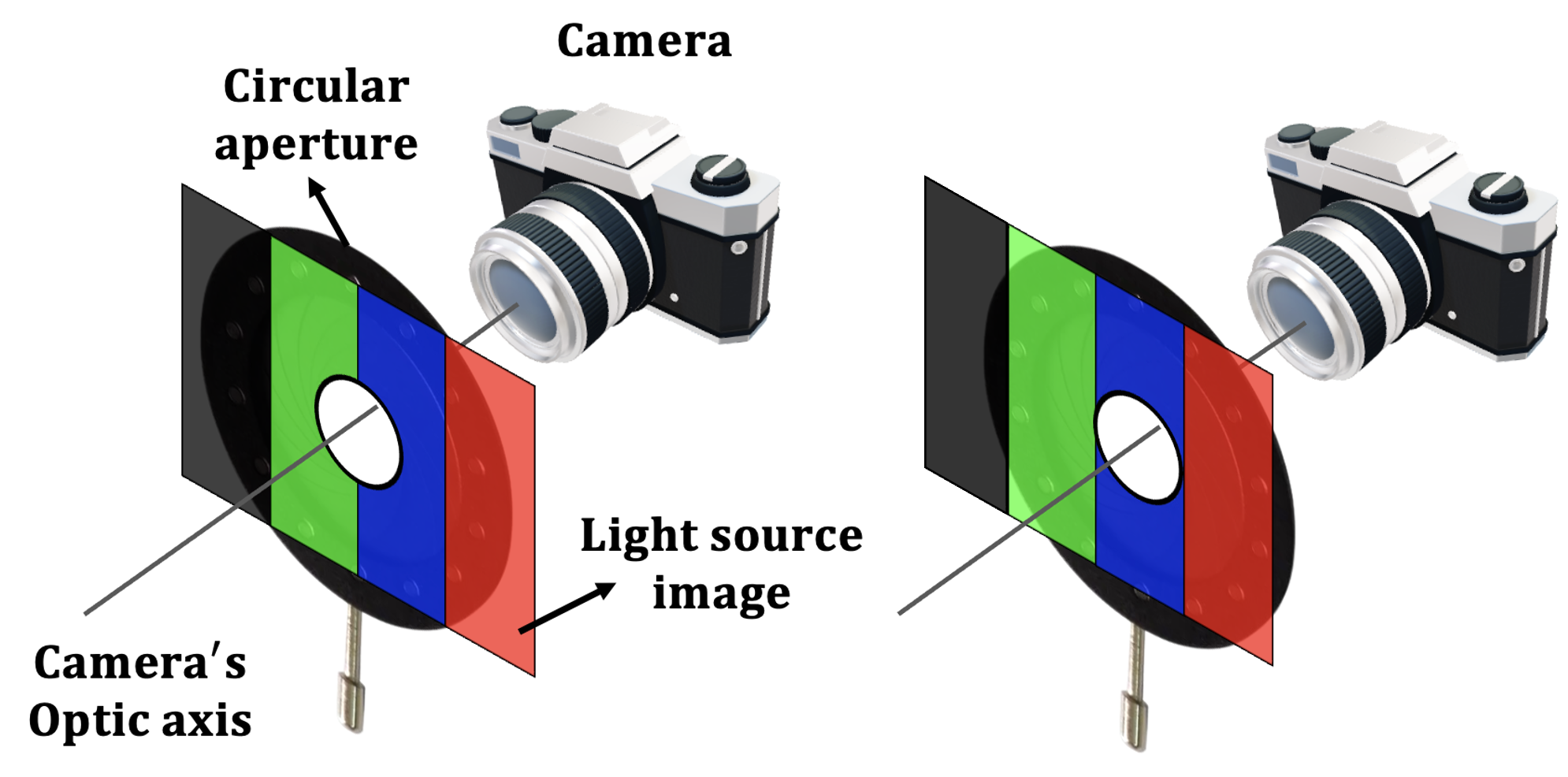}
        \caption{Light source image projection on aperture}
        \label{tradknifea}
    \end{subfigure}
    \hfill
    \begin{subfigure}[b]{0.46\textwidth}
        \centering
        \includegraphics[width=\textwidth]{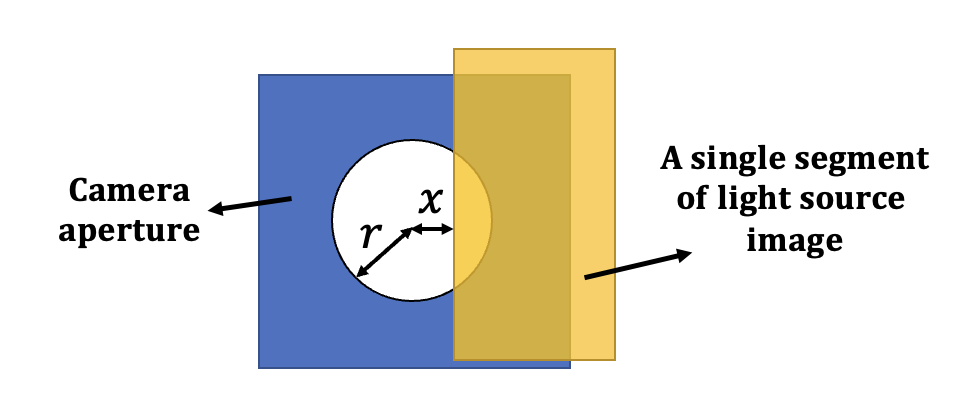}
        \caption{Single segment projection on aperture}
        \label{tradknifeb}
    \end{subfigure}
    \caption{The light source image projection on the aperture}
    \label{tradknife}
\end{figure}

To understand how gray-level variation arises with beam deviation, consider a single vertical segment of the light source imaged onto a circular aperture, with its width equal to the aperture diameter. When a portion of this segment is blocked by the aperture and the remaining part is allowed to pass, as illustrated in Fig.~\ref{tradknifeb}, a change in transmitted light intensity occurs for each respective channel. 

In an off-axis single-mirror Schlieren setup, let the luminance of the light source be \(L_o\), the mirror's focal length be \(f\), and \(A(x)\) denote the unobstructed area of the source image at the aperture. When no Schlieren object is present, the luminance at the image plane in the presence of the aperture is defined as \cite{settles2001schlieren, wakoya2022sensitivity}:

\begin{equation}
    L_b=\frac{L_o A(x_i)}{m^2(2f)^2}
\end{equation}
Where the area \(A(x_i)\) represents the unobstructed portion of the light source image after being partially blocked by the knife edge represented in Fig. \ref{tradknifeb}, and can be explicitly defined as:
\begin{equation}
    A(x_i)={r^2 \Big{[}cos^{-1} \frac{|x_i|}{r}-\frac{|x_i|}{r^2}(\sqrt{r^2-|x_i|^2})\Big] }
    \label{areaeq}
\end{equation}
Here, \(m\) denotes the magnification factor, defined as the ratio of the Schlieren image size to the physical test area. Propagation losses are neglected in this analysis. In the absense of the Schlieren object, \(L_b\) represents the background illumination level in the image. When the object is introduced, the weak image of the light source corresponding to each point on the Schlieren object is projected onto the aperture and experiences a slight positional shift due to light deflection at an angle \(\theta\) at the object plane.
\vspace{-0.3cm}

The modified coordinate becomes \(\vec{x_f}=\vec{x_i}+2f\vec{\theta}\), leading directly to a luminance change at the image plane given by:
\begin{equation}
    \Delta L=\frac{L_o |A(x_i)-A(x_f)|}{ m^2(2f)^2}
\end{equation}
The observed contrast (\(C\)) in Schlieren images originates from variations in luminance relative to the background illumination, which is observed in the image and defined as the ratio of luminance difference \(\Delta L\) to the background illumination \(L_b\). The Schlieren sensitivity/ contrast sensitivity, defined as \(S = \frac{dC}{d\theta}\), quantifies the rate at which contrast changes with the deflection angle \(\theta\), and can be expressed as:
\begin{equation}
    S=-\frac{4f \sqrt{r^2-(x_i+2f\theta)^2}}{r^2\bigg{[}cos^{-1}\frac{x_i}{r}-\frac{x_i}{r^2}\sqrt{r^2-x_i^2}\bigg{]}}
    \label{eq:sens}
\end{equation}
The above equation suggests that the condition \(x_i \rightarrow r\) yields the highest sensitivity, corresponding to the case of a dark background in the image.

\subsection{Calibration curves}

\begin{figure}[H]
    \centering\includegraphics[width=13.6cm]{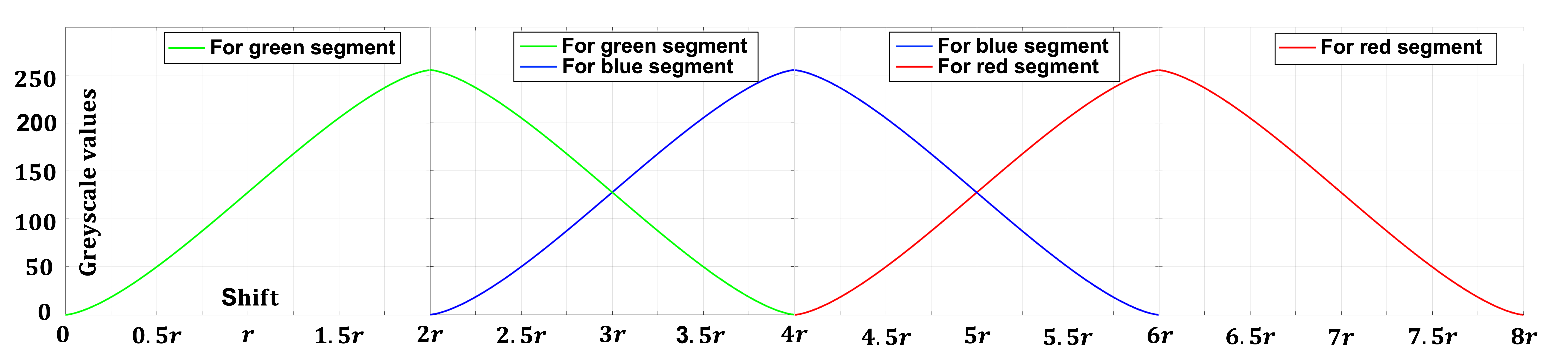}
    \caption{Calibration curves to calculate the image shift in New system.}
    \label{modelcurves}
\end{figure}

Local deflection in the Schlieren field, as described earlier, results in a lateral shift of the light source image on aperture plane.
While the contribution from a single segment may be partially or fully blocked (Fig.~\ref{tradknifeb}), additional light still passes through the aperture due to the extended nature of the segmented source (Fig.~\ref{tradknifea}). This produces image contrast either through grayscale variation within a colour channel or by introducing a new colour hue.
This contrast enables estimation of the tangential shift of the source image at the aperture plane. As illustrated in Fig.~\ref{tradknifea}, the aperture may transmit a single colour segment adjacent to a dark strip, or a combination of two colour segments, thereby defining the background composition in the Schlieren image. This relationship, along with Eq.~\ref{areaeq}, forms the basis for constructing calibration curves (e.g., Fig. \ref{modelcurves}) used in quantitative measurements.

\vspace{-0.3cm}

\begin{figure}[H]
    \centering
    \begin{subfigure}[b]{0.15\textwidth}
        \centering
        \includegraphics[width=\textwidth]{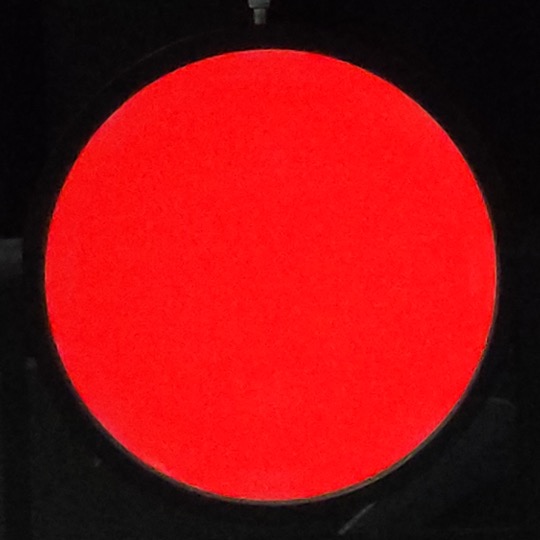}
        \caption{Red}
    \end{subfigure}
    \hfill
    \begin{subfigure}[b]{0.15\textwidth}
        \centering
        \includegraphics[width=\textwidth]{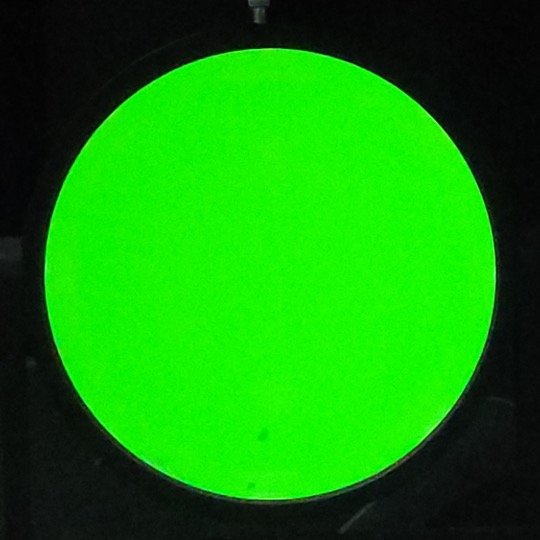}
        \caption{Green}
    \end{subfigure}
    \hfill
    \begin{subfigure}[b]{0.15\textwidth}
        \centering
        \includegraphics[width=\textwidth]{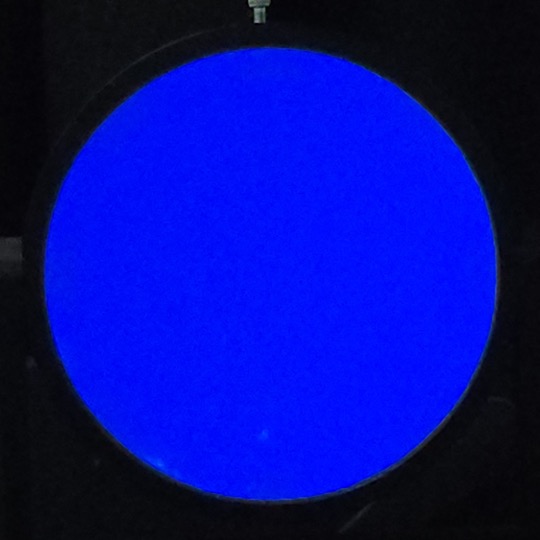}
        \caption{Blue}
    \end{subfigure}
    \hfill
    \begin{subfigure}[b]{0.15\textwidth}
        \centering
        \includegraphics[width=\textwidth]{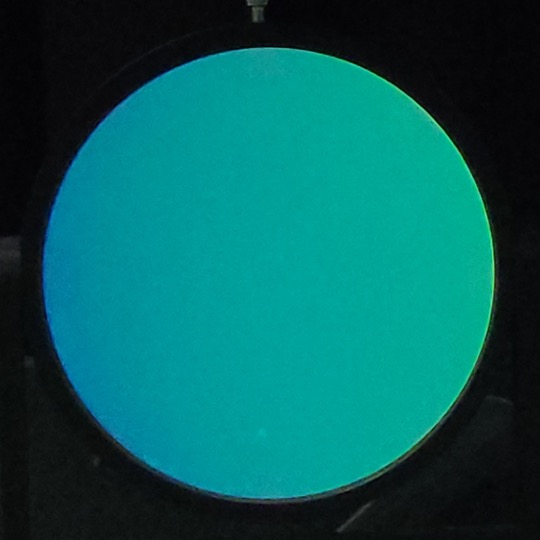}
        \caption{Blue-Green}
    \end{subfigure}
    \hfill
    \begin{subfigure}[b]{0.15\textwidth}
        \centering
        \includegraphics[width=\textwidth]{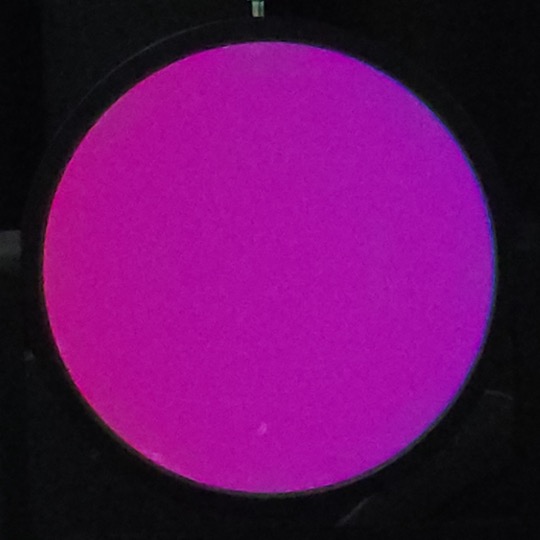}
        \caption{Red-Blue}
    \end{subfigure}
     \hfill
    \begin{subfigure}[b]{0.15\textwidth}
        \centering
        \includegraphics[width=\textwidth]{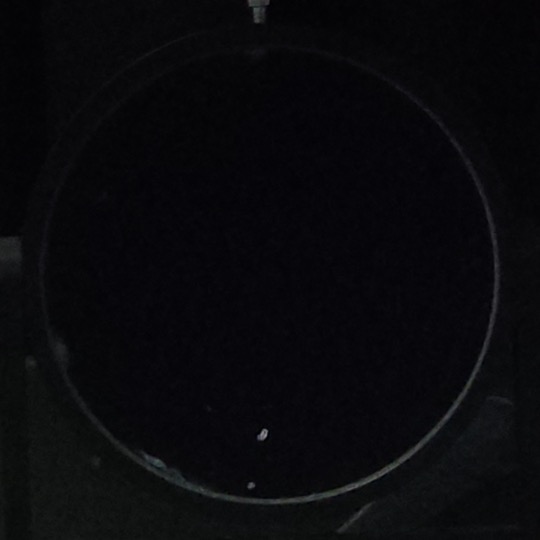}
        \caption{Black}
    \end{subfigure}
    \caption{Six different backgrounds in this scheme.}
    \label{sixbac}
\end{figure}

The acquired Schlieren image comprises two primary components: the background illumination and the main object. The image contrast varies with the refraction angle \(\theta\) and follows calibration curves that relate background intensity to shifts of the segmented RGB source at the aperture plane. In the proposed setup, the background can take one of six distinct colour combinations (Fig.~\ref{sixbac}): Red, Green, Blue, Black, Red-Blue, and Blue-Green.
Among these, Red and Green backgrounds are particularly versatile, as they can be fine-tuned to achieve the desired brightness without introducing additional colour hues. This allows us to obtain an image with a very low background luminance, providing precise control over the contrast between the background and the main object particularly achieving images with high sensitivity (Eq.~\ref{eq:sens}). Mixed backgrounds like Red-Blue and Blue-Green offer a graded response: weak to moderate refractive index gradients are represented by varying shades of the primary colours, while strong gradients appear as distinct third hues, corresponding to large deflection angles \(\theta\).
The Black (dark) background configuration is critical for applications such as shock wave detection, thermal plumes, and aerodynamic phenomena, which requires maximum contrast and sensitivity \cite{kleine1997dark}, see Eq.~\ref{eq:sens}. This is important as greatly enhances the visibility of subtle refractive index gradients and fine details.
\vspace{-0.3cm}
When the segment width matches the aperture diameter, distinct RGB combinations emerge until the shift reaches the periodic length of the segment set. To find the optimal aperture size, the total width of each segment set must be calculated based on the required measurement range. Since different Schlieren objects require different measurement ranges, a calibration experiment was conducted to determine the appropriate aperture and segment dimensions for a specific application.
\section{Calibration experiment}

To estimate the aperture size, we first need to find the required measuring range of the system for imaging the desired Schlieren object. In our experiment, we aim to visualize hot air around the combustion of butane gas. To find the necessary measuring range for this object, a calibration experiment was conducted using two small circular light sources of different colors, red and yellow, each with a diameter of \SI{2}{\milli\meter}.

\vspace{-0.5cm}
\begin{figure}[H]
    \centering\includegraphics[width=4.71cm]{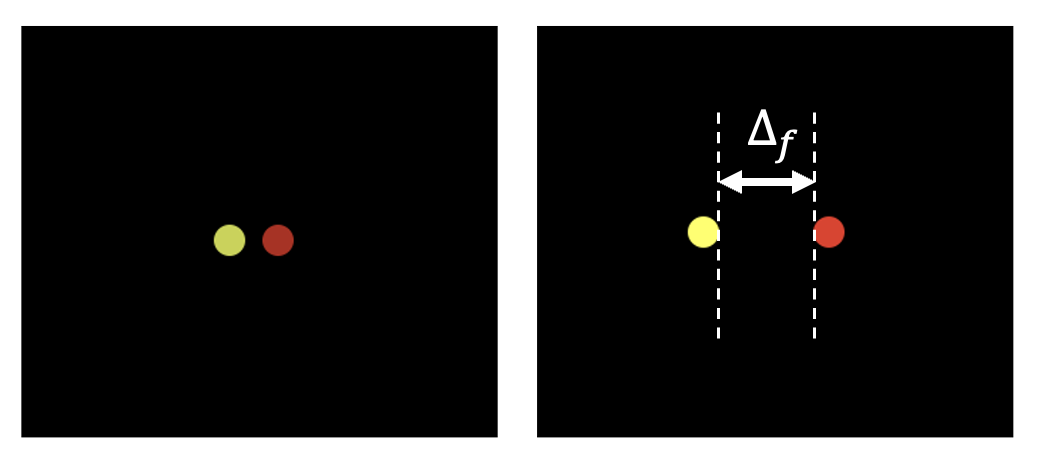}
    \caption{Calibration experiment light source}
    \label{lights1}
\end{figure}
One of the sources (red, in this experiment) is aligned with the optical axis of the camera, while another source (yellow) is offset by a small distance. In the absence of a Schlieren object, the captured image appears uniformly red. Upon introducing the object (e.g., butane combustion), refraction causes the yellow segment to contribute to the image due to the lateral shift of the light source image at the aperture plane. As the separation (\(s\)) between the red and yellow segments increases, a threshold is reached beyond which the maximum shift no longer allows yellow to enter the aperture, and it disappears from the image. To determine the required measuring range of the system, this segment separation is gradually increased until the significant appearance yellow hue is no longer observed. This critical distance corresponds to the width of one full set of four coloured segments and is used to calculate the aperture radius \(r\).
\begin{figure}[H]
    \centering\includegraphics[width=13.31cm]{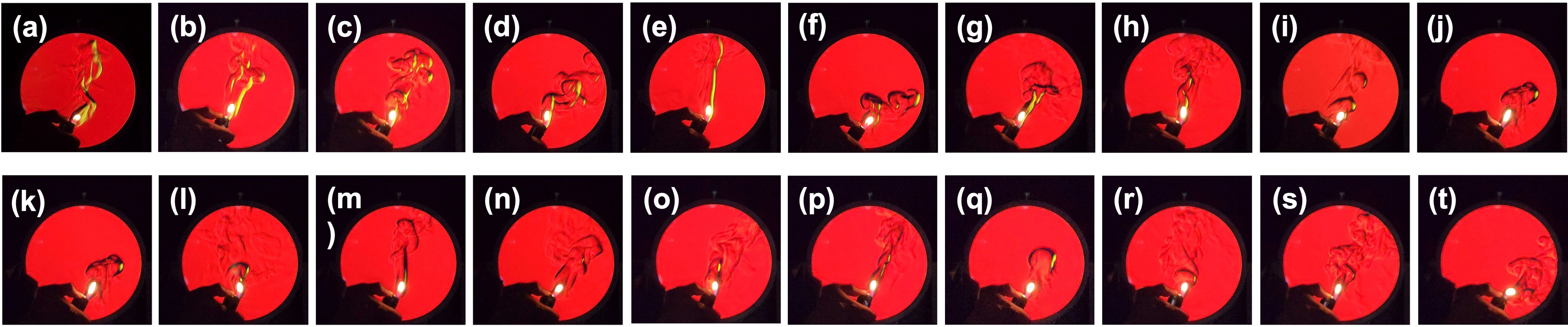}
    \caption{Schlieren images of butane gas combustion, with segment spacing \(s\) ranging from (a)~\SI{1.52}{\milli\meter} to (t)~\SI{6.04}{\milli\meter}.}
    \label{20fig1}
\end{figure}

The light source shown in Fig.~\ref{lights1} is used to captures 20 Schlieren images shown in Fig.~\ref{20fig1}. The initial separation between the yellow and red points is \SI{1.52}{\milli\meter} and a strong presence of yellow colour is observed in Fig.~\ref{20fig1}a. As the separation \(s\) between the two sources increases, the presence of yellow colour gradually diminishes (see Fig.~\ref{20fig1} (b--t)). Up to Fig.~\ref{20fig1}g, the yellow region remains clearly visible. However, in Fig.~\ref{20fig1}(h--s), the yellow becomes confined to a much smaller area, and by Fig.~\ref{20fig1}t, captured at a separation of \SI{6.04}{\milli\meter}, it disappears completely. This indicates that a measuring range corresponding to a shift of at least \SI{6.04}{\milli\meter} is required to fully visualise the hot air region generated by butane gas combustion.

The aperture radius is defined as the ratio of maximum separation between red and yellow sub-light sources \(\Delta_f\) and \(n\) as follows.
\vspace{-0.5cm}
\begin{equation}
    r = \frac{\Delta_f}{n} = \SI{1.51}{\milli\meter}
\end{equation}

After determining the aperture radius, we constructed the light source with each strip width of (\SI{3.02}{\milli\meter})
, equal to the aperture diameter.

\section{Reference system}
A reference system is designed using the traditional approach as shown in Fig.~\ref{tradknife1}. A square white light source with dimensions \(2r \times 2r\) was employed to match the size of a single segment in the proposed system (\SI{3.02}{\milli\meter}), ensuring comparable image sensitivity. A vertical knife edge is used as a cutoff to generate horizontal contrast in the image. To quantify the shift at the cutoff plane, a calibration curve was generated by capturing 40 images in the absence of the Schlieren object, with the knife edge incrementally displaced with step size of \SI{100}{\micro\meter}.
\vspace{-0.5cm}
\begin{figure}[H]
    \centering
    \begin{subfigure}[b]{0.865\textwidth}
        \centering
        \includegraphics[width=\textwidth]{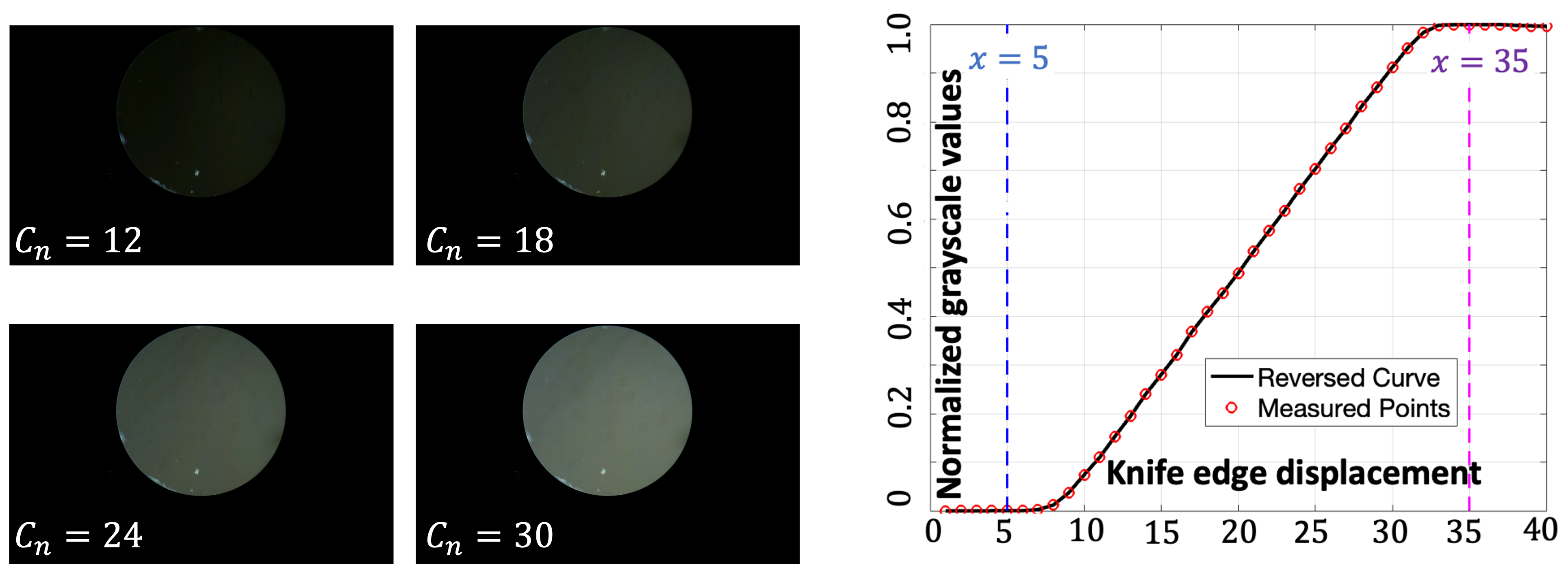}
        \caption{Calibration Schlieren Images with different knife edge positions}
        \label{20figa}
    \end{subfigure}
    \hfill

    \vskip\baselineskip

    \begin{subfigure}[b]{0.372\textwidth}
        \centering
        \includegraphics[width=\textwidth]{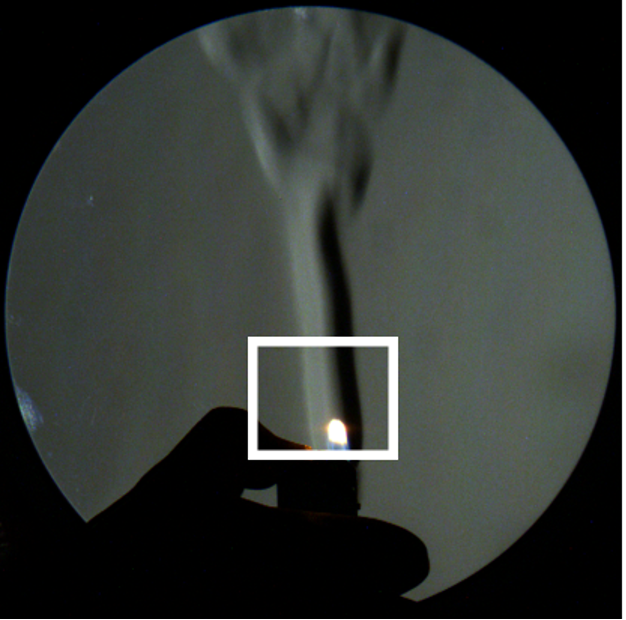}
        \caption{Captured image}
        \label{20figb}
    \end{subfigure}
    \begin{subfigure}[b]{0.48\textwidth}
        \centering
        \includegraphics[width=\textwidth]{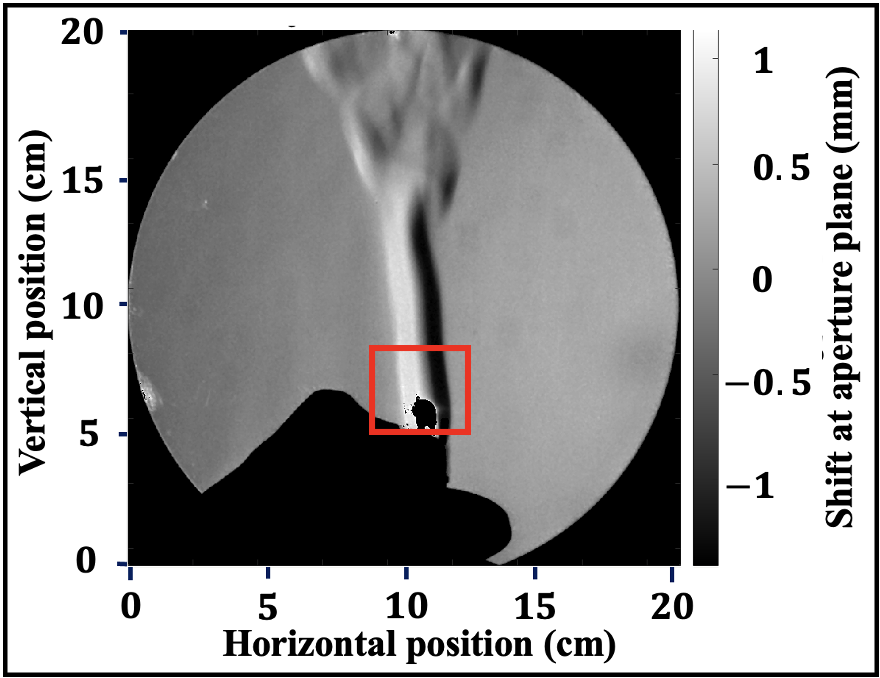}
\caption{Calculated image}
\label{20figc}
    \end{subfigure}
    \caption{Results obtained with reference system.}
    \label{20fig}
\end{figure}
The left 4 images in Figure \ref{20figa}, out of 40, are the Schlieren images captured in the absence of the Schlieren object. Each \(n^{th}\) captured image \((C_n)\) shows a uniform illumination for the respective position of the knife edge. Hence, a calibration curve can be constructed that relates the position of the knife edge to the normalized background illumination to quantify the reference image, shown in Fig.~\ref{20figb}.

\section{Experimental Results and Discussion}
In the experiment, hot air generated by butane gas combustion is used as the Schlieren object to verify the proposed system's ability to detect high light shifts.
In the reference system, a white light source with dimensions \(\SI{3.02}{\milli\meter} \times \SI{3.02}{\milli\meter}\) is used in the traditional system to match the sensitivity of the proposed system.
Fig.~\ref{20figb} shows the captured image using the reference system. A calibration curve is required to quantify this result by calculating weak source image shift corresponding to each element in the object.
The calibration curve is constructed from background illumination values obtained by capturing 40 Schlieren images (Fig.~\ref{20figa} without the Schlieren object at varying knife-edge positions. This curve is then used to calculate the shift corresponding to each pixel.
 The calculated image shown in Fig.~\ref{20figc} is constructed by first normalizing the grayscale values of Fig.~\ref{20figb} to a range \(0-1\) and then comparing them with the normalized grayscale values of the calibration curve. The corresponding knife-edge displacement values were determined, and the shift at the aperture plane was calculated by subtracting the knife-edge displacement of the background illumination (set to zero) from the knife edge displacement value for each pixel. The observed shift ranged from \SIrange{-1.406}{1.140}{\milli\meter} at the aperture plane, as shown in Fig.~\ref{20figc}.

\begin{figure}[H]
    \centering
    \begin{subfigure}[b]{0.372\textwidth}
        \centering
        \includegraphics[width=\textwidth]{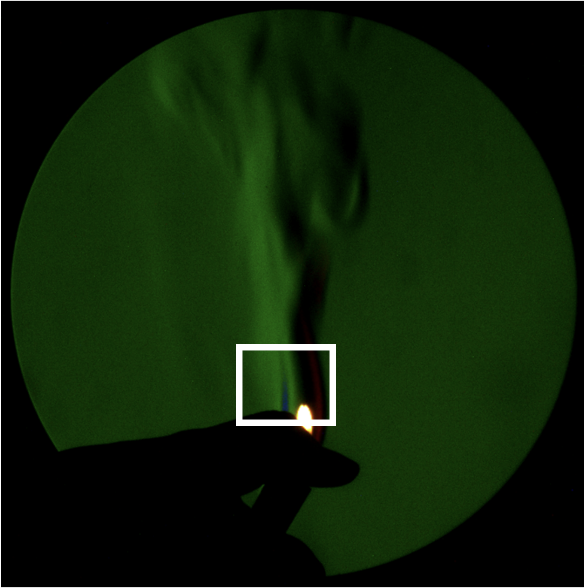}
        \caption{Observed image}
        \label{res2a}
    \end{subfigure}
    \begin{subfigure}[b]{0.49\textwidth}
        \centering
        \includegraphics[width=\textwidth]{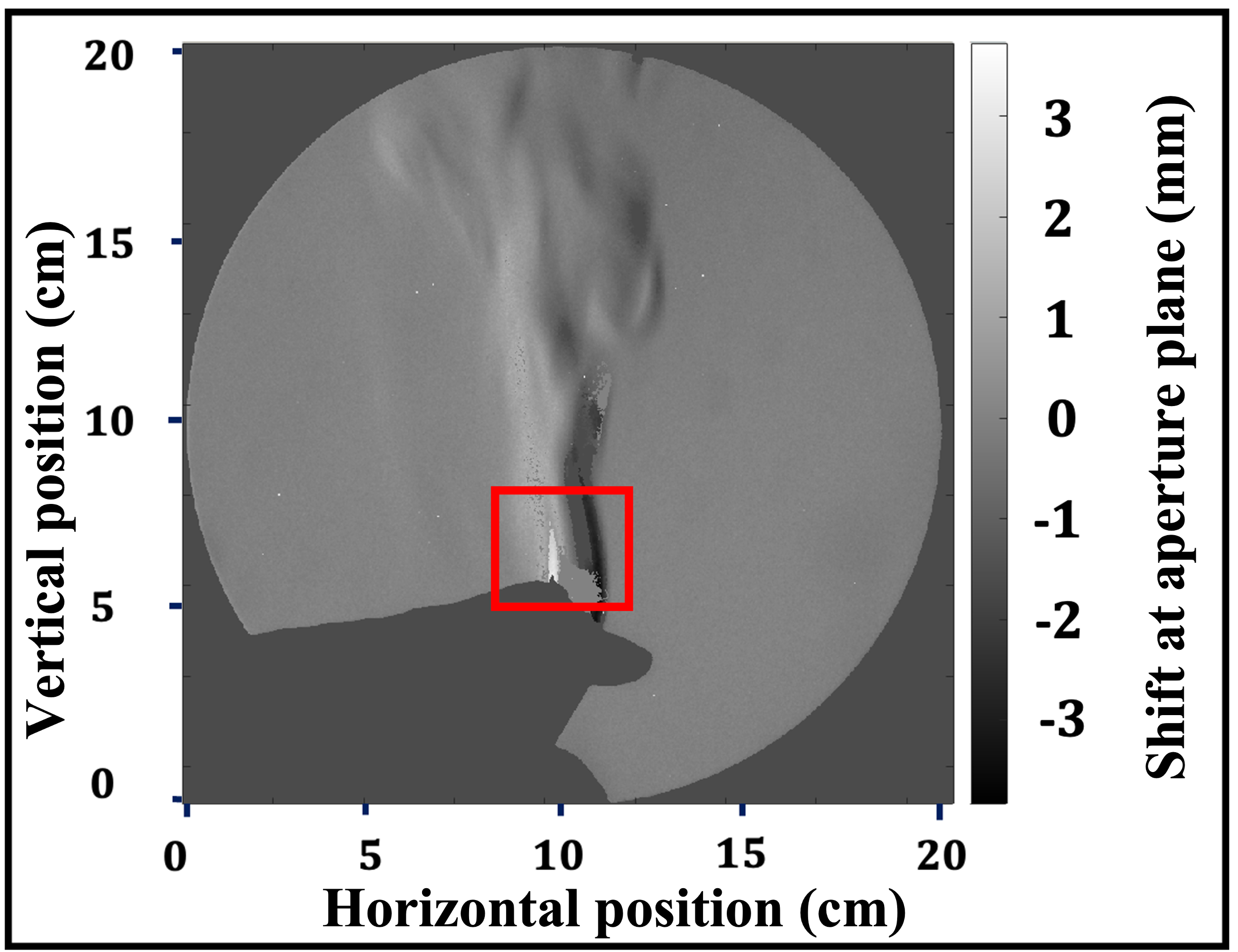}
        \caption{Calculated image}
        \label{res2b}
    \end{subfigure}
    \caption{Results obtained with the new system.}
    \label{res2}
\end{figure}

\begin{figure}[H]
    \centering
    \begin{subfigure}[b]{0.47\textwidth}
        \centering
        \includegraphics[width=\textwidth]{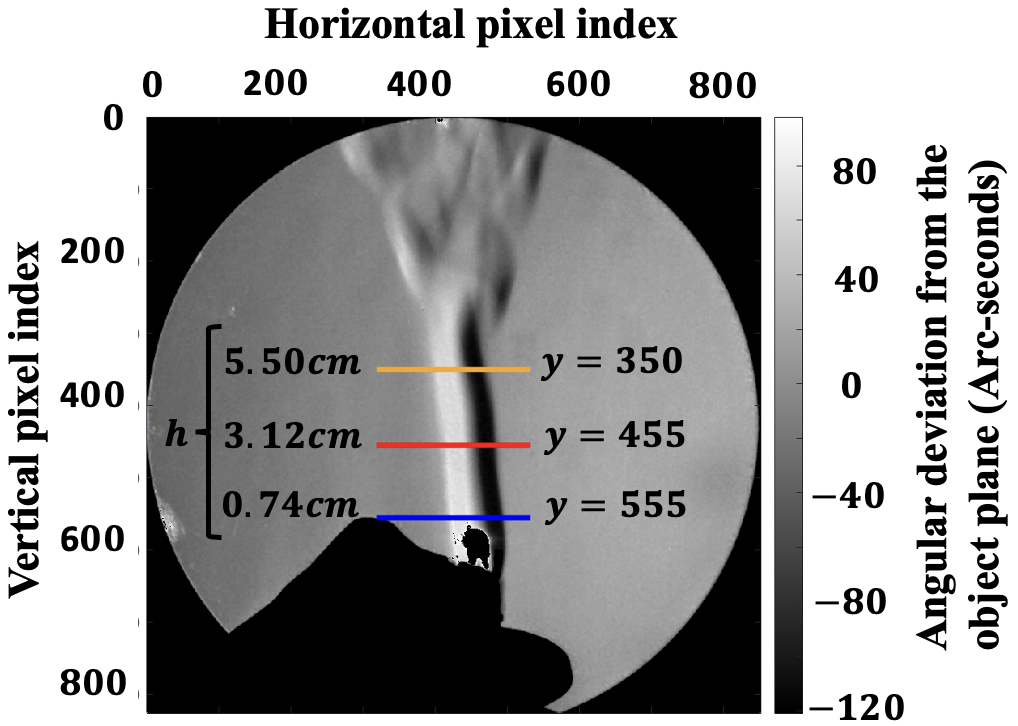}
        \caption{Schlieren image (reference system)}
        \label{linegraph1a}
    \end{subfigure}
    \hfill
    \begin{subfigure}[b]{0.47\textwidth}
        \centering
        \includegraphics[width=\textwidth]{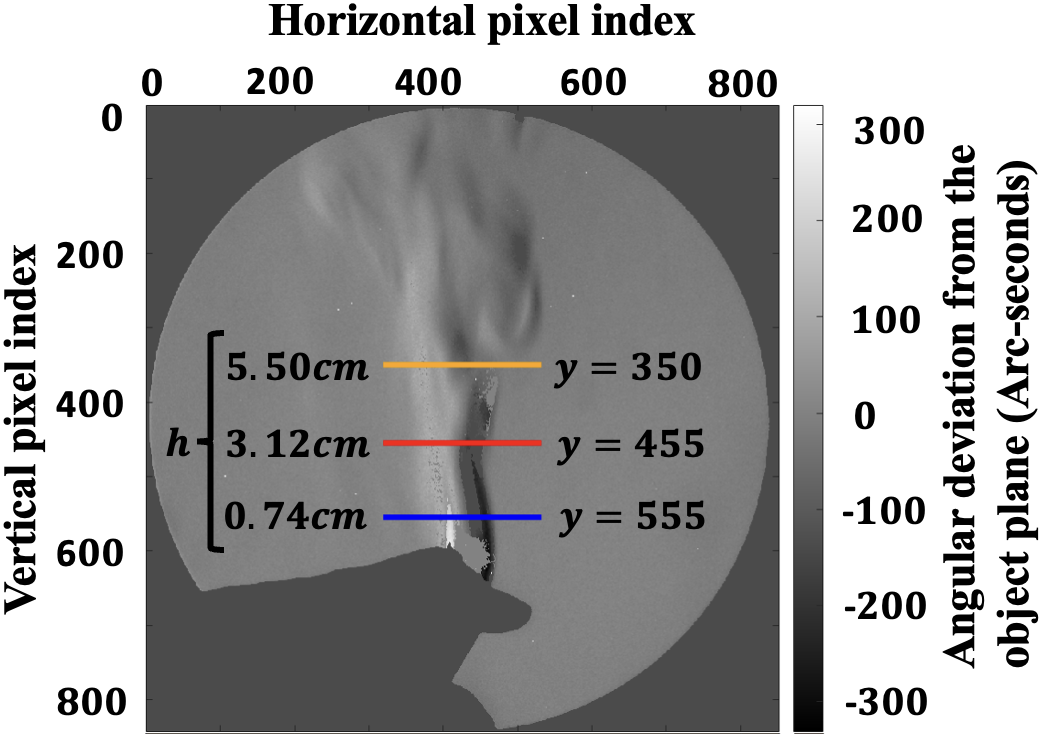}
        \caption{Schlieren image (new system)}
        \label{linegraph1b}
    \end{subfigure}
    \hfill
    \begin{subfigure}[b]{0.47\textwidth}
        \centering
        \includegraphics[width=\textwidth]{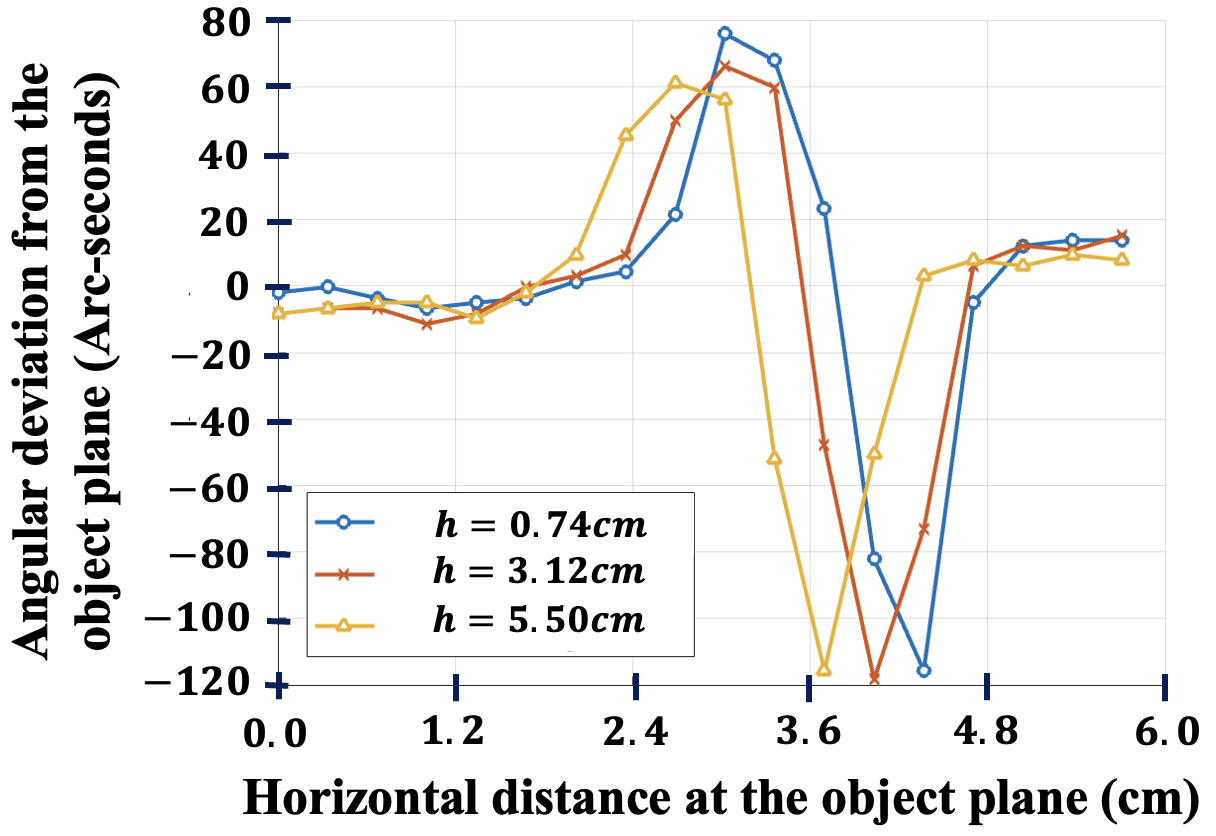}
        \caption{Line profile (reference system)}
        \label{linegraph1c}
    \end{subfigure}
    \hfill
    \begin{subfigure}[b]{0.47\textwidth}
        \centering
        \includegraphics[width=\textwidth]{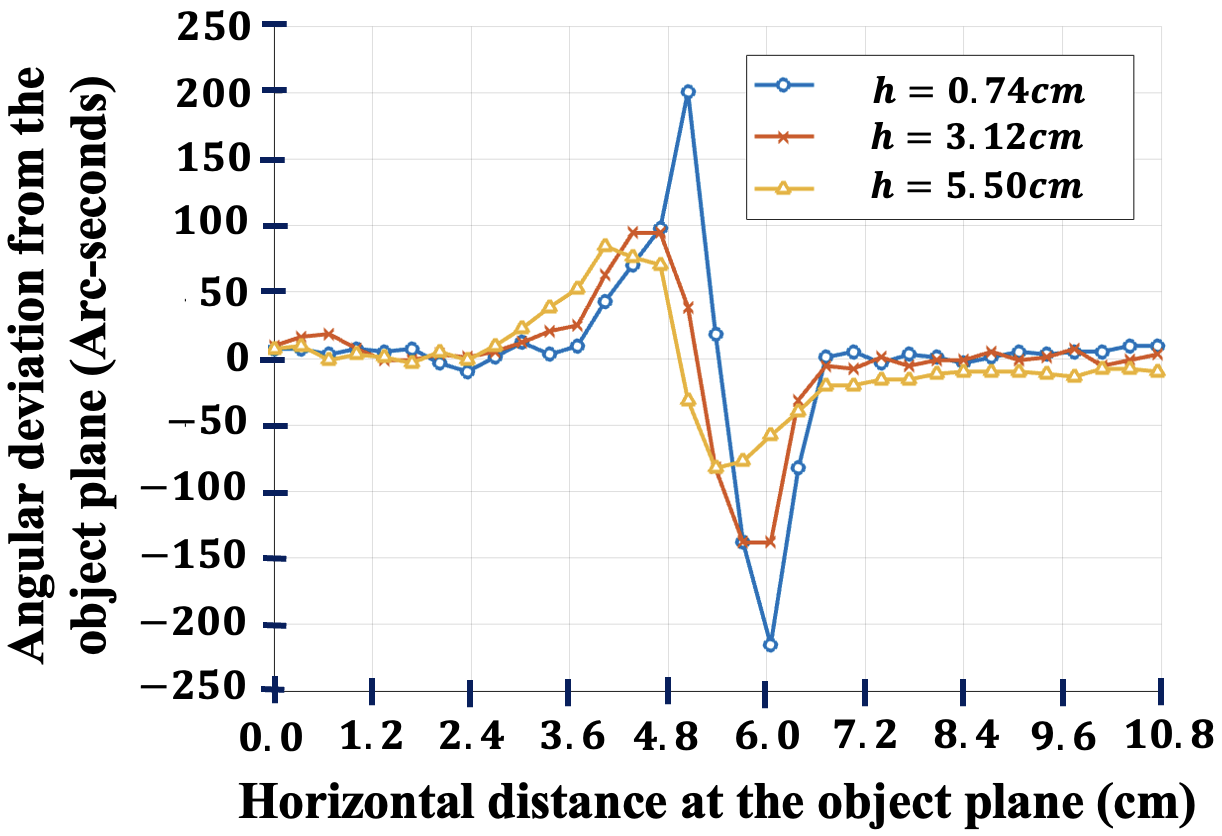}
        \caption{Line profile (new system)}
        \label{linegraph1d}
    \end{subfigure}
    \caption{Angular deviation line profile comparison.}
    \label{linegraph1}
\end{figure}

\begin{figure}[H]
    \centering
    \begin{subfigure}[b]{0.24\textwidth}
        \centering
        \includegraphics[width=\textwidth]{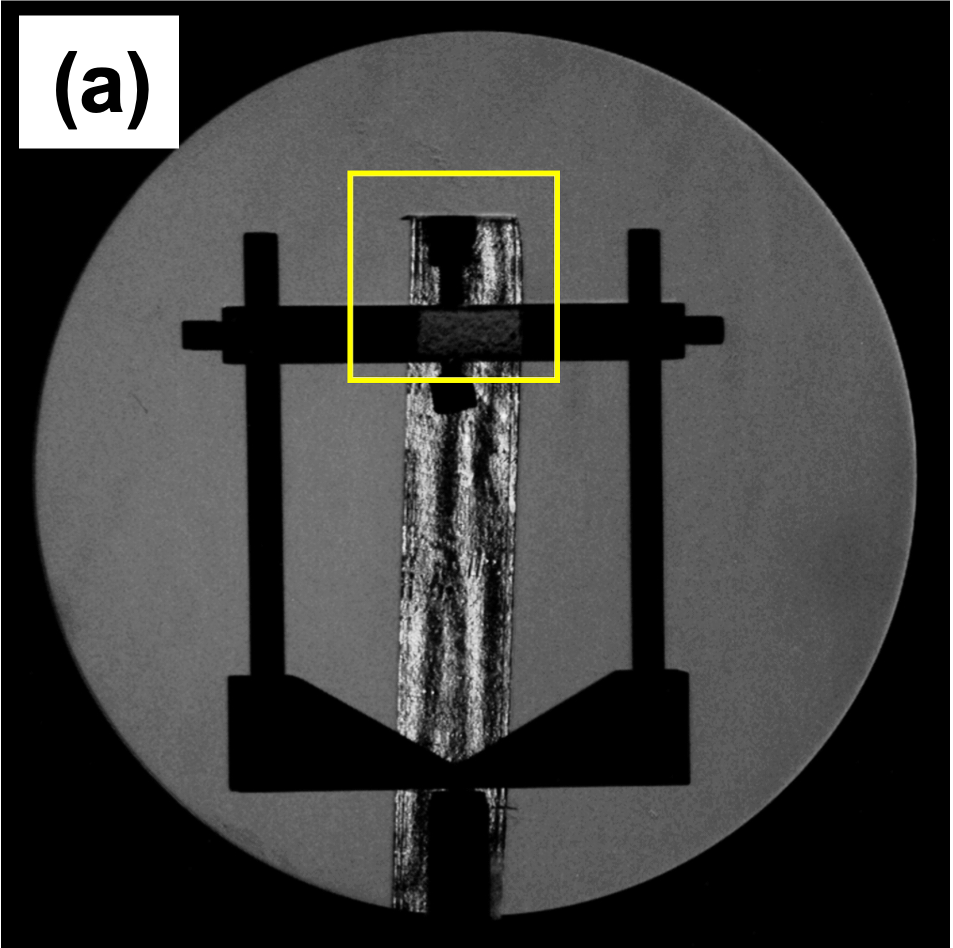}
    \end{subfigure}
    \hfill
    \begin{subfigure}[b]{0.24\textwidth}
        \centering
        \includegraphics[width=\textwidth]{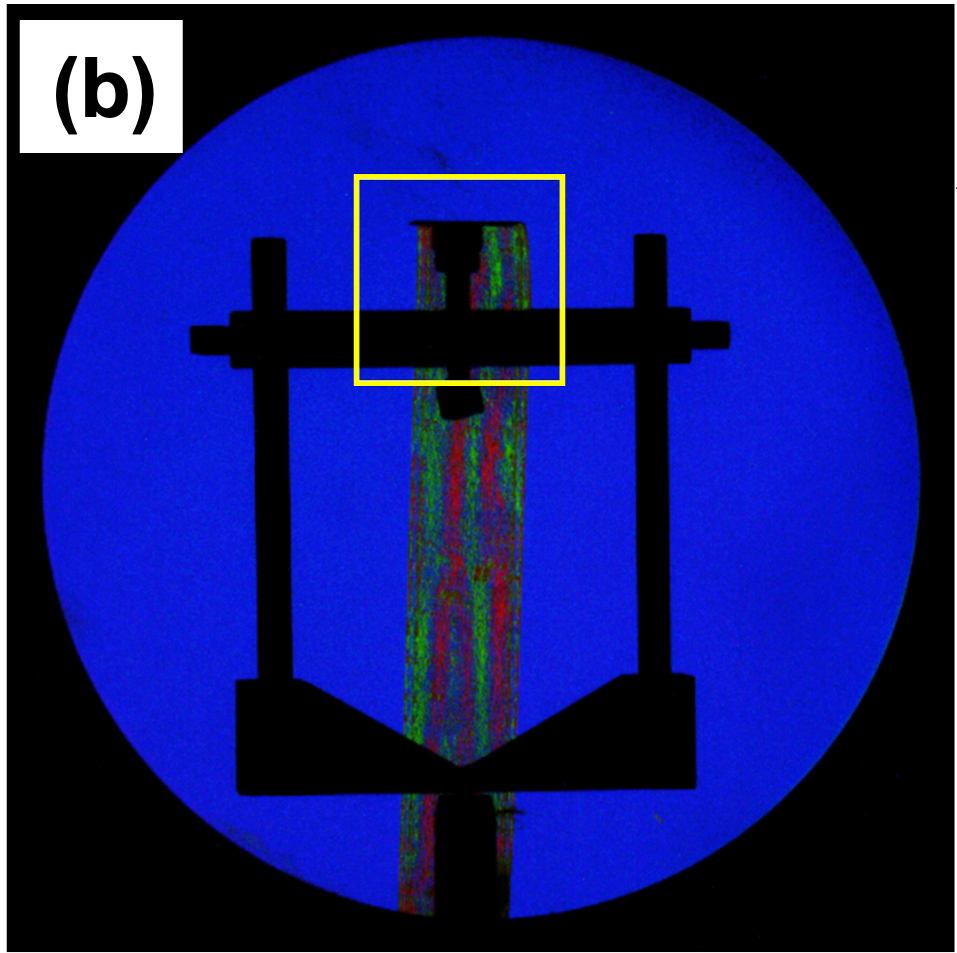}
    \end{subfigure}
    \hfill
    \begin{subfigure}[b]{0.24\textwidth}
        \centering
        \includegraphics[width=\textwidth]{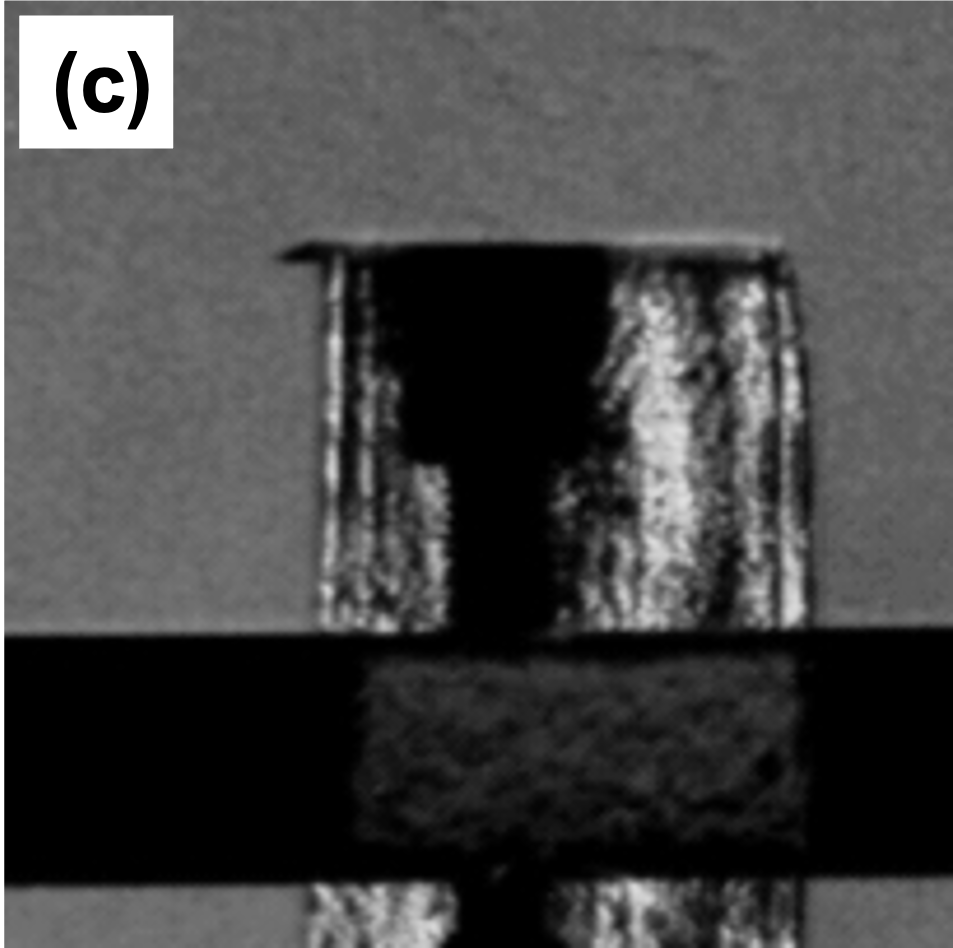}
    \end{subfigure}
    \hfill
    \begin{subfigure}[b]{0.24\textwidth}
        \centering
        \includegraphics[width=\textwidth]{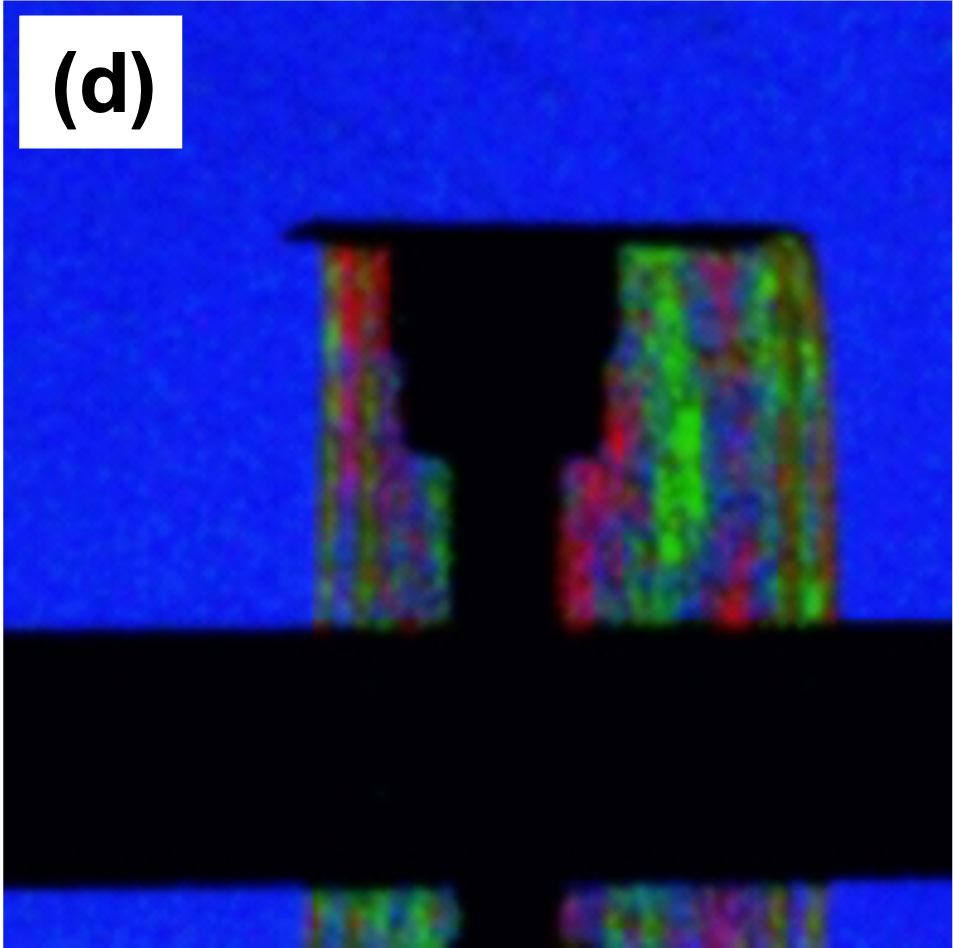}
    \end{subfigure}
    \caption{Schlieren images of a transparent adhesive tape. \textbf{(a)} Captured by the reference system. \textbf{(b)} Captured by the new system. \textbf{(c)} Zoom-in view of the image (a). \textbf{(d)} Zoom-in view of the image (b).}
    \label{res1m1}
\end{figure}

Fig.~\ref{res2a} shows the captured result with the newly proposed system, where the presence of green along with subtle blue and the red hues indicates strong light deviation due to a high refractive index change near the lighter plume, and shows that the gradients are not high enough to shift the deviated rays in Blue-Red region of calibration curves (Fig.~\ref{modelcurves}. However, in the reference image, no noticeable variation is observed in the same region in the image, even when examining different heights within the image. Using the new system calibration curves shown in Fig.~\ref{modelcurves}, we calculated the shift of the source image on the cutoff plane corresponding to each element of the object. The observed shift ranged from \SIrange{-3.845}{3.725}{\milli\meter} at the aperture plane which is 2.973 times larger than the range observed in the reference system. Theoretically, if all four regions of the calibration curves contributed, the expected range would be 4 times larger than the reference system. However, since one of these regions is not involved in the deviation, the theoretical expectation is reduced to 3 times. The experimental result closely matches this prediction, showing a measured increase of \textasciitilde 3 times. The area near the lighter plume exhibits a high shift, as shown in Fig.~\ref{res2b}. While this region shows a significant shift, the shift decreases when moving further upward. In this area, no substantial difference is observed between the reference system and the new system, which is due to the lower refractive index gradient.
\vspace{-0.3cm}
In Fig.~\ref{linegraph1}, three line profiles are plotted for heights \SIlist{0.74; 3.12; 5.50}{\centi\meter} from the combustion plume to compare the sensitivity of the proposed system to detect the angular shift of light rays with the reference system. At heights \SIlist{3.12; 5.50}{\centi\meter}, there is no noticeable difference between the measuring range of both systems. However, when observing this phenomenon near the combustion plume, at height \SI{0.74}{\centi\meter}, a significant difference in detection is observed between the two systems. The reference system could not detect the higher deviation, shown in Fig.~\ref{linegraph1a}. This is also evident in Fig.~\ref{20figc}, where minimal variation is observed up to a certain height above the combustion plume. On the other hand, in Fig.~\ref{linegraph1b}, the captured image with the proposed system displays a significantly high deviation, \SIrange{-215.289}{200.678}{\arcsecond} \text{ at height } \SI{0.74}{\centi\meter}
, which verify that the proposed system has become more sensitive to detect larger deviations.
\vspace{-0.3cm}
A separate experiment is conducted to qualitatively compare the results of the proposed and the reference system. In this experiment, a transparent adhesive tape is used as the Schlieren object. Two Schlieren images, one from each system, of the tape are captured to investigate its surface characteristics. The uneven surface of the tape likely causes slight variations in thickness or material density, which cause localized changes in the refractive index, resulting in differential refraction of light as it passes through the tape (Fig.~\ref{res1m1}). 
\vspace{-0.3cm}
The image reveals variations in grayscale contrast. Here, Fig.~\ref{res1m1}a shows the image captured by the reference system, which reveals variations in grayscale contrast. However, high-density gradient regions on the tape deviate light at a sufficiently large angle, resulting in the complete blockage of the deviated light by the knife edge. As a result, these gradients, and those exceeding this level, cannot be distinguished and appear black in the image. Additionally, an opaque stand in the setup also appears black in the image, as it blocks the light entirely. This overlap in black regions caused by both high-density gradients and the opaque stand creates ambiguity, making it challenging to clearly distinguish the boundaries of the opaque object from the Schlieren image regions (Fig.~\ref{res1m1}c). The same image is captured using the new proposed system (Fig.~\ref{res1m1}b). Investigating carefully the same small area in this image (Fig.~\ref{res1m1}d), we find that the boundaries of the opaque are visible and differentiable to the Schlieren object unlike the Fig.~\ref{res1m1}c. Additionally, we can see more details due to coded colours in the image.

\begin{figure}[H]
    \centering
    \begin{subfigure}[b]{1.01\textwidth}
        \centering
        \includegraphics[width=\textwidth]{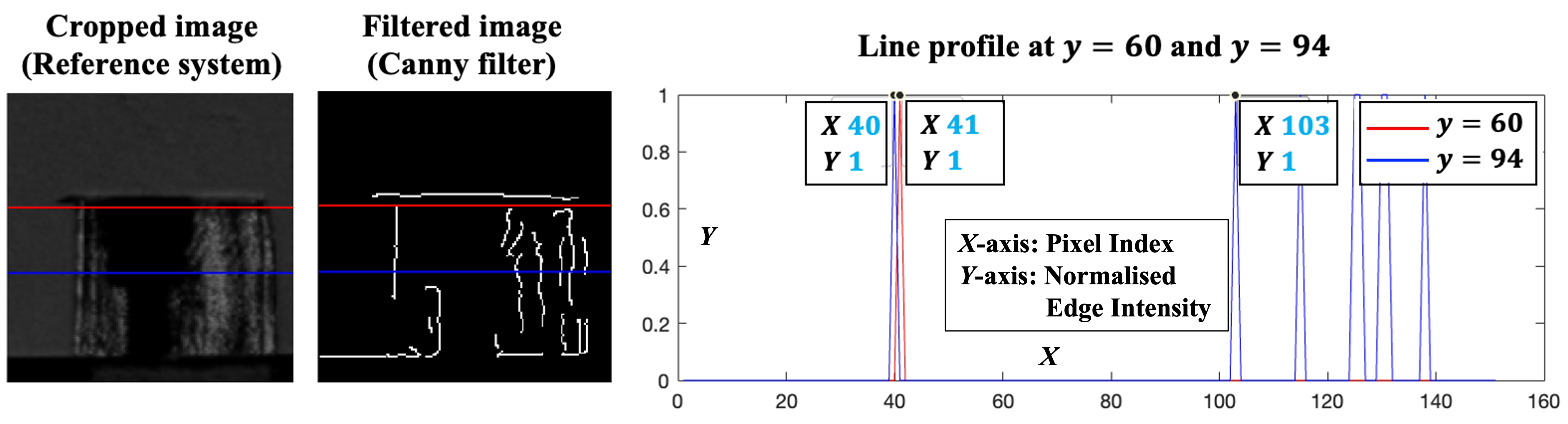}
        \caption{Image analysis captured by the reference system}
        \label{restape1a}
    \end{subfigure}
    \hfill
    \begin{subfigure}[b]{1.01\textwidth}
        \centering
        \includegraphics[width=\textwidth]{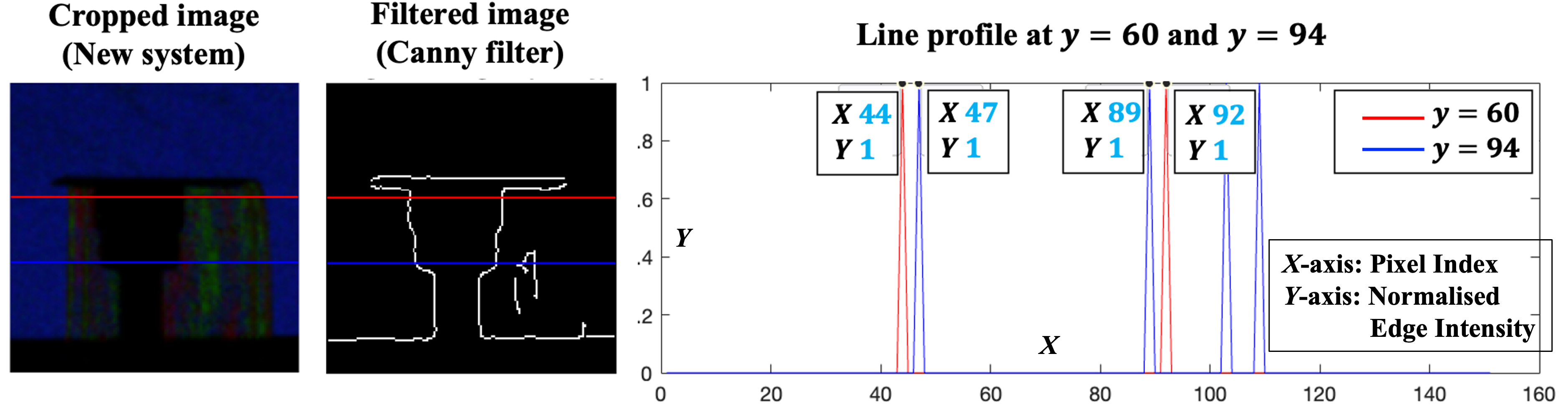}
        \caption{Image analysis captured by the proposed system}
        \label{restape1b}
    \end{subfigure}
    \hfill
    \begin{subfigure}[b]{1.01\textwidth}
        \centering
        \includegraphics[width=\textwidth]{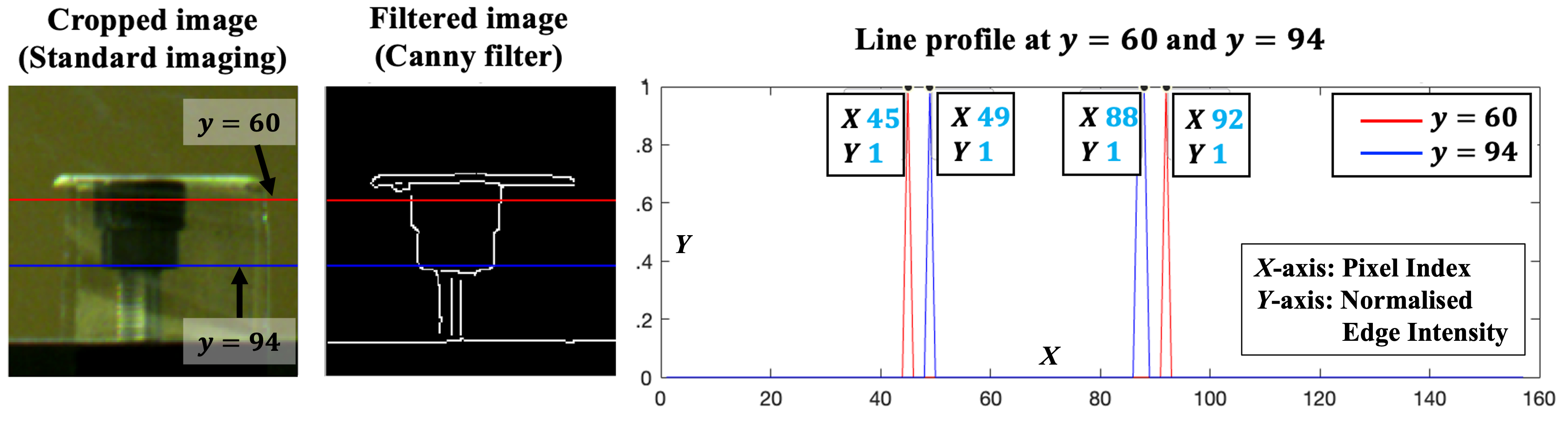}
        \caption{Analysis of the real image}
        \label{restape1c}
    \end{subfigure}
    \caption{Edge analysis.}
    \label{restape1}
\end{figure}

To accurately compare the results, both Schlieren images were processed using a Canny edge detection filter with low and high threshold values of \(0.4\) and \(0.62\), respectively, as shown in Fig.~\ref{restape1a} and \ref{restape1b}. The filtered images highlight the edges of the opaque stand mount, allowing for a clearer comparison of the sensitivity of each system to provide better features in the image. However, to verify the accuracy of the captured image/edges, an image of the stand and the adhesive tape was captured without the Schlieren system. The Canny edge detection filter was then applied to identify the true edges of the opaque stand.
\vspace{-0.3cm}
The comparison between Fig.~\ref{restape1a} and \ref{restape1c} demonstrates that the reference system did not accurately capture the edges. However, the image taken by the proposed system (Fig.~\ref{restape1b}) is very close to the true edges. To evaluate edge detection accurately, two line profiles were extracted from each edge image at y-pixel values of \(60\) and \(94\). At \(y=60\), in Fig.~\ref{restape1c}, edges were observed at x-pixel positions of \(45\) and \(88\). The proposed Schlieren system provided comparable results, where edges are detected at \(x=44\) and \(x=89\). However, a 5-pixel shift was observed in the reference system edge image in the first x value, while the second edge was not detected. At \(y=94\), the true edges were observed at x-pixel positions \(49\) and \(94\). In Fig.~\ref{restape1b}, these edges are observed at \(x=47\) and \(x=92\), closely matching the true edges. In contrast, the reference system's edge image showed a notable shift, with edges detected at \(x=41\) and \(x=103\), deviating significantly from the position of the true edges.

\begin{figure}[H]
    \centering\includegraphics[width=13.3cm]{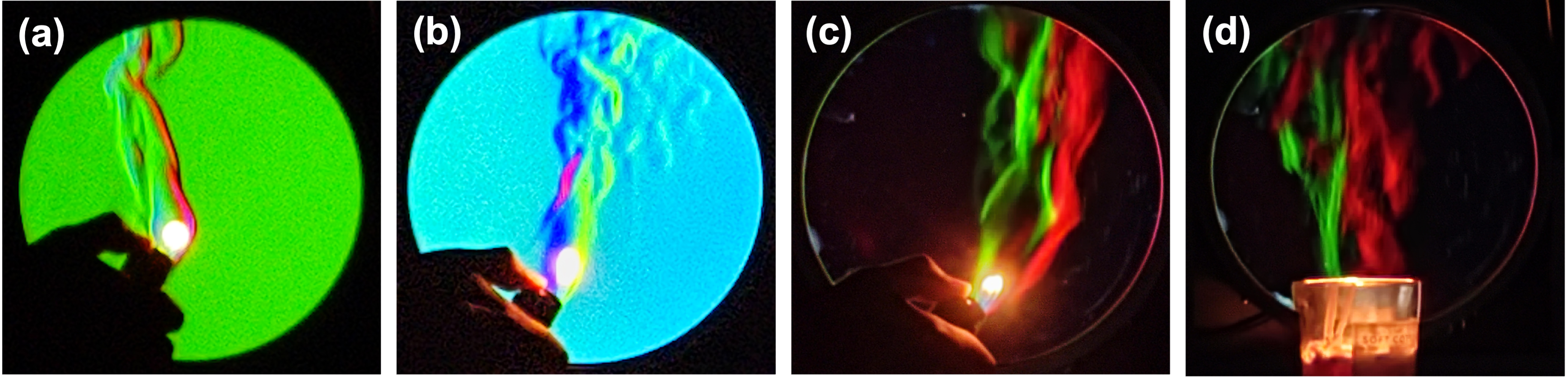}
    \caption{Schlieren images of hot air from butane combustion (a--c) and a candle plume (d), captured using an external-cutoff-free setup with a mobile camera: (a) green, (b) green-blue, and (c--d) dark backgrounds.}
    \label{res1m}
\end{figure}

Furthermore, to make the system completely knife edge/external cut-off free, the external aperture is completely eliminated to utilise the camera's own aperture. A mobile phone camera was chosen as an imaging device for this purpose for its compactness and portability, which, while offering less precision than specialized scientific cameras, provides a practical and convenient solution for environments where traditional imaging systems may be too bulky or difficult to use. The first two images in Fig.~\ref{res1m} show Schlieren images in two different bright-coloured backgrounds. However, the third and fourth images were captured using a dark background by aligning the black segment image of the light source with the camera. The first three images visualise hot air generated by the combustion of butane gas using a lighter; however, in the fourth image, the hot air is generated using a candle plume.
\vspace{-0.3cm}

Therefore, the proposed Schlieren setup, which leverages an extended segmented light source in conjunction with a circular aperture, demonstrates significant advantages.
Most notably, it extends the measuring range of the Schlieren system, its primary strength, and simplifies alignment, making it particularly effective in complex experimental environments.
This novel arrangement has proven to have major benefits over the traditional system.
Utilising these new components, we observed greater light deviation due to the system's extended measuring range compared to the traditional approach, while maintaining the same sensitivity. This enhanced range was further validated qualitatively using a transparent adhesive tape as a Schlieren object.

Additionally, replacing the knife edge with a circular aperture not only removes the alignment complexity but also avoids the sensitivity fluctuations associated with traditional approach~\cite{schmidt2015quantitative}.
Due to the self-aligning nature of the circular aperture, the system eliminates the need for external cutoffs (e.g., knife edge/aperture/filters) and instead directly utilizes the camera's built-in aperture.
A mobile phone can also be used as an imaging device; its aperture diameter typically ranges between \(2\,\mathrm{mm}\) and \(4\,\mathrm{mm}\), which is well-suited for creating a self-aligned setup for many Schlieren applications.
The system's low cost and compatibility with widely available smartphones make it an attractive option for commercialization and educational use. Furthermore, modern smartphones often feature built-in image processing capabilities which makes it an excellent choice for qualitative analysis.
However, for scientific applications requiring raw, unprocessed data and greater control over aperture settings, dedicated cameras with adjustable apertures are more appropriate.
Although the circular aperture offers notable advantages over the knife edge when integrated with the proposed segmented light source, it still encounters limitations at very small aperture sizes. Diffraction effects begin to manifest, which can blur the image and degrade overall quality. Additionally, a smaller aperture reduces light transmission, leading to lower exposure and potentially diminished sensitivity. This limitation can be mitigated by employing a sufficiently intense light source. Future work should explore optimization of the aperture design or alternative filtering mechanisms to minimize diffraction-related issues. Another challenge is the current use of only four segments in the light source. Increasing the number of segments could further extend the system's measuring range or enhancing the sensitivity for a fixed measurement range. Since the number of segments has a linear relationship with sensitivity, higher segmentation may enhance both resolution and contrast. Further research should investigate advanced segmentation strategies to maximize performance.

Furthermore, the system can be further modified to measure the temperature across different ranges in a single observation, e.g., a temperature at the boundary of a flame is significantly higher than its body temperature. Moreover, the potential to miniaturise this system for portable or in situ measurements is an exciting possibility. The self-aligning nature and reduced complexity of the setup could lead to the development of compact Schlieren devices suitable for conditions where space and alignment constraints are critical factors.
\section{Conclusion}
In this work, we propose a novel knife-edge-free Schlieren imaging system that uses a new light source and is integrated with a circular aperture. The proposed system offers a large measuring range of the instrument and easier alignment compared to the traditional system. In the experimental investigation, we have achieved a \textasciitilde 3 times larger measurement range, which also produced clearer edge details compared to a traditional system, avoiding underexposure and overexposure, and qualitative Schlieren images of hot air using the portable, knife-edge free design with a mobile phone camera. Although certain challenges remain, such as diffraction at smaller apertures, the segmentation of the light source proposed here provides a novel approach to increase the sensitivity and the measuring range of the imaging system, which enables new possibilities for its application in various scientific and industrial domains such as optical flow visualisation and quantitative analysis.
\bibliography{references}
\begin{appendix}
\section{Appendix}
When a light source with slit area A and luminance Lo illuminates a mirror's surface placed at a distance D, then by the inverse square law, the illuminance L on the mirror is defined as:
\begin{equation}
    L_A=\frac{L_o A}{D^2}
\end{equation}

In an off-axis Schlieren system, the light source illuminates the mirror placed at a distance \(2f\), which then reflects and forms the source image on the aperture plane. Assuming negligible propagation losses, the same illuminance \(L_A\) reaches the test area. In the absence of any cutoff at the light source image plane, the Schlieren image also maintains this illuminance level, but reduced by a factor of \(m^2\), where \(m\) is the magnification factor defined as the ratio of the Schlieren image size to the test area size. Additionally, since the mirror is located at a distance \(2f\), we substitute \(D = 2f\). Hence, the Schlieren image illuminance is given by:

\begin{equation}
    L_A=\frac{L_o A}{m^2 (2f)^2}
\end{equation}

The illuminance in the image scales with the square of the magnification factor (\(m^2\)), as it corresponds to areal magnification, where the image area increases proportionally to the square of the linear magnification. Since the light source image is formed at the aperture plane, an aperture placed at this location allows a portion of the light with area \(A(x_i)\) to pass through. Consequently, the background illumination in the Schlieren image is given by:

\begin{equation}
    L_b=\frac{L_o A(x_i)}{m^2 (2f)^2}
\end{equation}

Here, the area \(A(x_i)\) corresponds to the unobstructed portion of the light source image and is defined in Eq.~\eqref{areaeq}.
\end{appendix}
\end{document}